\documentclass[aps,prd,a4paper,showpacs,twocolumn]{revtex4-1}
\usepackage{amsmath, amsthm, amssymb}
\usepackage{mathtools}
\usepackage{nicefrac} 
\usepackage{epsfig}
\usepackage{graphicx}
\usepackage[dvipsnames]{xcolor}
\usepackage{mathrsfs} 
\usepackage{amssymb} 
\usepackage{amsmath} 
\usepackage{xspace} 

\usepackage[bookmarks, breaklinks, colorlinks,urlcolor=black, citecolor=red, 
linkcolor=blue]{hyperref}
\usepackage{float}
\def\dsp{\displaystyle}
\def\nn {\nonumber}

\def\gev{\ensuremath{\mathrm{Ge\kern -0.1em V}}}
\def \eff{{\text{eff }}}
\def \kstar{{K^{\!*}}}

\def\korkst{{K^{\!(\!*\!)}}}
\def\rk {R_K}
\def\rkst {R_{K^{\!*}}}
\def\rkrkst {R_{K^{\!(\!*\!)}}}
\def\rl {R_\Lambda}

\allowdisplaybreaks

\begin{document}
\title{Lepton mass effects and angular observables in $\Lambda_b \to 
\Lambda (\to p \pi) \ell^+\ell^-$ 
}  
\author{Shibasis Roy}\email{shibasisr@imsc.res.in} 
\affiliation{The Institute of Mathematical
	Sciences, Taramani, Chennai 600113, India \\ and \\ Homi Bhabha National 	
	Institute Training School Complex, \\ Anushakti Nagar, Mumbai 400085, 
	India} 
\author{Ria Sain}\email{riasain@imsc.res.in} 
\affiliation{The Institute of Mathematical
	Sciences, Taramani, Chennai 600113, India \\ and \\ Homi Bhabha National 
	Institute Training School Complex, \\ Anushakti Nagar, Mumbai 400085, 
	India} 
\author{Rahul Sinha}\email{sinha@imsc.res.in}
\affiliation{The Institute of Mathematical
	Sciences, Taramani, Chennai 600113, India \\ and \\ Homi Bhabha National 
	Institute Training School Complex, \\ Anushakti Nagar, Mumbai 400085, 
	India}

\date{\today}
\begin{abstract}
The flavor changing rare decay $B\to\kstar(\to K\pi)\ell^+\ell^-$ is one of the
most studied modes due to its sensitivity to physics beyond the standard model
and several discrepancies have come to light among the plethora of observables
that are measured. In this paper we revisit the analogous
baryonic decay mode $\Lambda_{b}\rightarrow \Lambda (\to p\pi) \ell^{+}\ell^{-}$
and we present a complete set of ten angular observables that can be measured
using this decay mode. Our calculations are done retaining the finite lepton
mass so that the signal of lepton non-universality observed in
$B\to\kstar\ell^+\ell^-$ can be corroborated by the corresponding baryonic decay
mode.  We show that due to the parity violating nature of the subsequent
$\Lambda\to p\pi$ decay there exists at least one angular asymmetry that is non-vanishing in the large recoil
limit unlike the case in $B\to\kstar\ell^+\ell^-$ decay mode, making it
particularly sensitive to new physics that violates lepton flavor universality.
\end{abstract}
\maketitle 
\section{Introduction} 

It is well known that the rare decay $B\to \kstar\ell^+\ell^-$ involves a  $b\to
s$ flavor changing loop induced transition at the quark level making it
sensitive to physics beyond the standard model
(SM)~\cite{Das:2012kz,Mandal:2014kma, Mandal:2015bsa,
Kruger:2005ep,Altmannshofer:2008dz,
Bobeth:2008ij,Egede:2008uy,Bobeth:2010wg,Becirevic:2011bp,Bobeth:2012vn,Gratrex:2015hna,
Grinstein:2004vb,Altmannshofer:2013foa,Hiller:2013cza}. The nature of this decay provides one with a
significant number of observables, many of which have been recently measured
\cite{Aaij:2013qta, Chatrchyan:2013cda} to a great deal of accuracy. There are
several discrepancies observed when compared to the SM predictions, among 
these, 
$\rkrkst$, the ratio of the differential decay rate $d(B\rightarrow
\korkst\ell^{+}\ell^{-})/dq^2$, for $\ell=\mu$ and $e$, has generated a great 
deal of interest. The deviation of $\rkrkst$ from the expected value in the SM 
imply a 
challenge to the idea of
lepton universality~\cite{other-signals} within the SM and points towards a possible evidence of new
physics (NP). Naturally the question arises whether we can observe a similar
deviation in other decay modes that capture this non-universal behavior of the
leptons. This will go a long way in establishing lepton non-universality on firm
footing. Here we reexamine the analogous baryonic decay of $\Lambda_{b}$ to
$\Lambda$ and a lepton-antilepton pair, where the $\Lambda$-baryon further
decays to proton $p^+$ and a pion $\pi^{-}$ as already discussed by various
authors Ref.~\cite{Gutsche:2013oea,Gutsche:2013pp,
Boer:2014kda,Mott:2011cx,Kumar:2015tnz,
Leitner:2006nb,Chen:2001zc,Huang:1998ek,Chen:2001ki,Chen:2001sj,Aliev:2002hj,
Aliev:2002nv,Aliev:2002tr,Aliev:2004yf,Giri:2005yt,Aliev:2006xd,
 Aliev:2006gv,Zolfagharpour:2007eh,Aslam:2008hp,Aliev:2010uy,Wang:2008sm,Wang:2009hra,
 Sahoo:2009zz,Wang:2015ndk,Gutsche:2017wag,Detmold:2016pkz,Meinel:2016grj,Faustov:2017wbh,Blake:2017une}.
 The 
underlying quark level $b\rightarrow s \ell^{+}\ell^{-}$ transition for 
$\Lambda_{b}\rightarrow \Lambda \ell^{+}\ell^{-}$ decay is the same as in the 
well studied $B\rightarrow K^{(*)}\ell^{+}\ell^{-}$ decay,  making it an ideal 
candidate to study in depth.

Before we study the consequences of lepton non-universality in baryonic decay 
$\Lambda_{b}\rightarrow
\Lambda(\to p\pi) \ell^{+}\ell^{-}$, we recall, that $\rkrkst$ is defined 
\cite{Aaij:2017vbb} within a given range of
the dilepton mass squared $q^2_{\rm min}$ to $q^2_{\rm max}$ as,
\begin{equation}
\label{eq:rkrkst}
\rkrkst= \frac{\dsp \int_{q^2_{\rm min}}^{q^2_{\rm max}}\frac{d 
\Gamma(B\rightarrow \korkst\mu^+\mu^-)
}{dq^2}dq^2}{\dsp\int_{q^2_{\rm 
min}}^{q^2_{\rm max}} \frac{d 
\Gamma(B \rightarrow\korkst e^+e^-)}{dq^2}dq^2}
\end{equation} 
The measured $\rk$ and $\rkst$, lie systematically 
below the SM expectations \cite{bifani,1406.6482}:
\begin{align}
&{\rk}{(q^2 \in [1:6]\, \gev^2)} = \dsp 0.745^{+0.090}_{-0.074}\pm 0.036 
,\nn \\
&\rkst(q^2 \in [0.045:1.1]\, \gev^2) =\dsp {0.660}^{+0.110}_{-0.070} \pm 
0.024,\nn \\
&\rkst(q^2 \in [1.1:6]\, \gev^2) = \dsp 0.685 ^{+0.113}_{-0.069} \pm 0.047 \nn.
\end{align}
In the SM both $\rk$ and $\rkst$ are predicted to be virtually  
indistinguishable from unity~\cite{sm-pred} for $(q^2 \in [1:6]\, \gev^2)$, 
whereas $\rkst\sim 0.9$ for $q^2 \in [0.045:1.1]\, \gev^2$ owing to a finite 
$m_\mu$. The measurements correspond to a $2.6\sigma$, $2.1\sigma$ and 
$2.4\sigma$
shortfalls from the SM expectations respectively. 

It is obvious that an observable $R_\Lambda$ can be proposed in the same spirit 
as 
$\rkrkst$ for the 
corresponding baryonic decay $\Lambda_b\to \Lambda \ell^+\ell^-$ as,
\begin{equation}
\label{eq:rl}
R_\Lambda=\frac{\dsp\int_{q^2_{\rm min}}^{q^2_{\rm max}} \frac{d 
\Gamma(\Lambda_{b} 
\rightarrow \Lambda\mu^+\mu^-)}{dq^2}dq^2}{\dsp\int_{q^2_{\rm 
min}}^{q^2_{\rm 
max}} 
\frac{d \Gamma(\Lambda_{b} \rightarrow \Lambda e^+e^-)}{dq^2}dq^2}
\end{equation}
One should expect the lepton mass effect to play a significant role on $\rl$ in
the low-$q^2$ region, just as in the case of $\rkrkst$. The discrepancy between
the SM expectation and the experimentally observed value of $\rkrkst$ was largest in
the low-$q^2$ region. The observation of a similar discrepancy in $R_{\Lambda}$
is therefore necessary to substantiate the idea of lepton non-universality in
FCNC processes since such an observation cannot  be restricted to the
celebrated $B\rightarrow K^{(*)}\ell^{+}\ell^{-}$ alone. In order to disentangle
the new physics contribution that may manifest as lepton non-universality, one
must take into account the SM contribution to $R_\Lambda$ including the effect
of finite leptons mass~\cite{Kadeer:2005aq,Gutsche:2015mxa}. We therefore derive the expression for $\rl$ without any
approximation.

Another salient feature of the $\Lambda_{b}\rightarrow \Lambda(\rightarrow
p^{+}\pi^{-}) \ell^{+}\ell^{-}$ decay, is the wealth of information carried by
the angular observables expressed in terms of the angular asymmetries, of which
the forward-backward asymmetry in the hadron angle $\theta_{\Lambda}$ is of
particular interest. We show that due to the parity violating nature of the
$\Lambda\to p\pi$ decay angular asymmetry are non-vanishing in the large recoil
or low-$q^2$ limit unlike the case in $B\to\kstar(\to K\pi)\ell^+\ell^-$, making it
particularly sensitive to new physics that violates lepton flavor
universality. This follows since ratios of hadronic forward-backward asymmetry
for $\ell=\mu$ and $\ell=e$, in the low-$q^2$ region, is a ratio of two finite
quantities for  $\Lambda_{b}\rightarrow \Lambda(\rightarrow p^{+}\pi^{-})
\ell^{+}\ell^{-}$ decay. It may be recollected that all asymmetries for
$B\to\kstar\ell^+\ell^-$ decay mode vanish in the low-$q^2$ region~\cite{Hiller:2013cza}.

Our paper is arranged in the following way; in Sec.~\ref{sec:angular-distribution} we derive the complete angular 
distribution consisting ten angular observables retaining all the helicities and lepton mass. 
Sec.~\ref{sec:helicity amplitudes} is devoted to the calculation of hadronic 
helicity amplitudes in terms of known parameters namely the Wilson coefficients 
and form factors. In Sec.~\ref{Decay rate and angular observables} the decay 
rate and angular asymmetries are written in terms of the helicity amplitudes. 
We also define observables that are free from hadronic uncertainties. Finally 
we conclude how these observables can play an important role in pinning down 
lepton-universality violating new physics.

\section{The decay of $\Lambda_{b}\rightarrow \Lambda(\rightarrow p^{+}\pi^{-}) 
\ell^{+}\ell^{-}$ }
\label{sec:angular-distribution}
The process $\Lambda_{b}\rightarrow \Lambda(\rightarrow p\pi^{-})+j_{\text{eff}}(\rightarrow \ell^{+}\ell^{-})$ can be thought of as a sequential decay where it is assumed that the daughter $\Lambda$-baryon is onshell and subsequently decays resonantly. This enables one to write down a joint angular decay distribution ~\cite{Chung:1971ri, Bialas:1992ny, Richman:1984gh, Korner:2014bca} which is described fully by four independent kinematic variables: the dilepton invariant mass squared $q^2$, the polar angles $\theta_{l}$, $\theta_{\Lambda}$ and the azimuthal angle $\phi$ defined by the decay products in their respective centre-of-mass (CM) frames. At this point we would like to clarify that in our convention we have chosen $\theta_{l}$ to be the angle between the lepton ($\ell^{-}$) and the  flight direction of the $j_{\text{eff}}$-system, $\theta_{\Lambda}$ to be the angle between the nucleon ($p$) and the $\Lambda$ flight direction and $\phi$ to be the angle between the two decay planes. 
The angular distribution involves the helicity amplitudes 
$H_{\lambda_{\Lambda},\lambda_j}(J)$ for the decay 
$\Lambda_{b}\rightarrow\Lambda + j_{\text{eff}}$,  $h^{a}_{\lambda_1,\lambda_2}$ for 
the decay $j_{\text{eff}} \rightarrow \ell^+ \ell^-$ and $h_{\lambda_p,0}$ for the 
decay $\Lambda \rightarrow p+\pi^-$. The joint angular distribution for a 
unpolarized $\Lambda_{b}$ decay is given by,
\begin{widetext}
\begin{eqnarray}
\label{eq:ang dist first}
K(q^2,\theta_l,\theta_\Lambda,\phi)=&\sum_{J,J',M_i=\pm
\frac{1}{2},M_i'=\frac{1}{2},\lambda_j,\lambda_{j}',a,a',\lambda_{\Lambda},
\lambda_{\Lambda}',\lambda_p,\lambda_1,\lambda_2}H_{\lambda_{\Lambda},
\lambda_j}^{a}(J) H_{\lambda_{\Lambda'},\lambda_j'}^{a'*}(J') \rho_{M_i,M_i'}
\nonumber \\ &\mathcal{D}^{\frac{1}{2}}_{M_i,\lambda_{\Lambda}-\lambda_j}(0,0,0)
\mathcal{D}^{*\frac{1}{2}}_{M_i',\lambda_{\Lambda}'-\lambda_j'}(0,0,0)
\delta_{JJ'}
h^{a}_{\lambda_1,\lambda_2}(J)h^{a'*}_{\lambda_1,\lambda_2}(J')\nonumber \\&
\mathcal{D}^{J}_{\lambda_j,\lambda_1-\lambda_2}(0,\theta_l,0)
\mathcal{D}^{*J'}_{\lambda_j',\lambda_1-\lambda_2}(0,\theta_l,0)
h_{\lambda_p,0}h_{\lambda_p,0}^*\\&
\mathcal{D}^{\frac{1}{2}}_{\lambda_{\Lambda},\lambda_p}
(-\phi,\theta_{\Lambda},\phi)\mathcal{D}^{*\frac{1}{2}}_{\lambda_{\Lambda'},
\lambda_p}(-\phi,\theta_{\Lambda},\phi) \nonumber
\end{eqnarray}
\end{widetext}
The polarization density matrix of $\Lambda_b$, $ \rho_{M_i,M_i'}$ in 
Eqn~\eqref{eq:ang dist first} is a 
hermitean 2$\times$2 matrix, with $\text{Tr}(\rho)$ = 1. $\rho_{++}$ and $\rho_{--}$
represent the probability that the initial state $\Lambda_b$  has
$M_i=\frac{1}{2}$ and $M_i=-\frac{1}{2}$ respectively. For an unpolarized sample
of $\Lambda_{b}$-baryon, $ \rho_{M_i,M_i'}=\frac{1}{2}\delta_{M_i,M_i'}$. In the
rest frame of $\Lambda_b$-baryon, the daughter $\Lambda$-baryon and
$j_{\text{eff}}$ fly back to back and without loss of generality it can be
assumed that the motion of $\Lambda$ and $j_{\text{eff}}$ is along the $z$-axis.
This reduces the first two Wigner's D functions to Kronecker delta functions
$\delta_{M_i,\lambda_{\Lambda}-\lambda_j}$ and
$\delta_{M_i',\lambda_{\Lambda}'-\lambda_j'}$ respectively, where $M_i,\
M_i'=\pm \frac{1}{2}$. After summing over $M_i,\ M_i'$ we are left with a
Kronecker delta
$\delta_{\lambda_{\Lambda}'-\lambda_j',\lambda_{\Lambda}-\lambda_j}$ which
signifies the fact that we considered the decay of an unpolarized $\Lambda_{b}$.
We also observe that $\vert\lambda_{\Lambda}'-\lambda_j'\vert= \vert
\lambda_{\Lambda}-\lambda_j\vert=\frac{1}{2}$ as the initial $\Lambda_{b}$ is
spin-1/2. This condition further restricts the values $\lambda_{\Lambda},\
\lambda_j$ can take and the fact has been already taken into account while calculating $K(q^2,\theta_l,\theta_\Lambda,\phi)$.
The choice of possible values for $\lambda_{\Lambda}$ and $\lambda_{j}$ are 
depicted in Table~\ref{Table:1}.

The hadronic helicity amplitudes $H_{\lambda_{\Lambda},\lambda_j}^{a}(J)$
contain all the information of the $\Lambda_{b}\rightarrow
\Lambda+j_{\text{eff}}$ transition in terms of the relevant form factors and
Wilson coefficients parametrizing the underlying $b\rightarrow s
\ell^{+}\ell^{-}$ process, explained in detail in Sec~\ref{sec:helicity amplitudes}. In case of a spin-$\frac{1}{2}$ $\Lambda_{b}$ baryon
decaying to an intermediate onshell spin-$\frac{1}{2}$ $\Lambda$ baryon there are four
hardonic helicitiy amplitudes           
$H_{\lambda_{\Lambda},\lambda_j}^{a}(J)$, where the index `$a$' denotes whether
the hadronic helicity amplitudes multiply the lepton vector current ($a=1$), or
the axial vector current ($a=2$). The label ($J$) takes the values ($J=0$) with
$\lambda_{j}=t$ and ($J=1$) with $\lambda_{j}=\pm 1,0$ for scalar and vector
parts of the effctive current $j_{\text{eff}}$ respectively. 
\begin{center}
\begin{table}
\begin{tabular}{c c c}
\hline
\hline
$\lambda_{\Lambda}$&$\lambda_j$&$M_i$\\
\hline
$1/2$&1&$-1/2$\\

$1/2$&0&$1/2$\\

$-1/2$&-1&$1/2$\\

$-1/2$&0&$-1/2$\\
\hline
\hline
\end{tabular}
\caption{The possible values of $\lambda_{\Lambda}$ and $\lambda_{j}$}
\label{Table:1}
\end{table}
\end{center}

Let us also discuss here the helicity amplitudes 
$h^{a}_{\lambda_{j};\lambda_{1},\lambda_{2}}$ appearing 
in Eq.~\eqref{eq:ang dist first} describing the process 
$j_{\text{eff}}\rightarrow \ell^+  
\ell^-$, 
where $\lambda_{j}=\lambda_{1}-\lambda_{2}$. Explicitly,
\begin{eqnarray} a=1\text{(V)}: \quad
h^{1}_{\lambda_{j};\lambda_{1},\lambda_{2}}(J)=\bar{u}_{1}(\lambda_1)\gamma_{\mu}v_{2}(\lambda_2)\epsilon^{\mu}(\lambda_j),\quad \nonumber \\  \\
a=2\text{(A)}: \quad
h^{2}_{\lambda_{j};\lambda_{1},\lambda_{2}}(J)=\bar{u}_{1}(\lambda_1)\gamma_{\mu}\gamma_5
 v_{2}(\lambda_2)\epsilon^{\mu}(\lambda_j). \nonumber
\end{eqnarray}
These helicity amplitudes are evaluated in the ($\ell^+\ell^-$) CM frame with 
$\ell^-$ defined in the $-z$ direction. The label  ($J$) is the same as defined previously and takes the 
values ($J=0$) with $\lambda_{j}=0(t)$ and ($J=1$) with $\lambda_{j}=\pm 1,0$ 
for scalar and vector parts of the effective current $j_{\text{eff}}$ 
respectively. The 
leptonic helicity amplitudes are calculated and given below: 
\begin{eqnarray}
h^{1}_{t;\frac{1}{2},\frac{1}{2}}(J=0)&= 0, \nonumber\\
h^{2}_{t;\frac{1}{2},\frac{1}{2}}(J=0)&= 2m_{\ell}, \nonumber \\
h^{1}_{0;\frac{1}{2},\frac{1}{2}}(J=1)&= 2m_{\ell}, \nonumber\\
h^{2}_{0;\frac{1}{2},\frac{1}{2}}(J=1)&= 0, \nonumber\\
h^{1}_{1;\frac{1}{2},-\frac{1}{2}}(J=1)&= -\sqrt{2q^2}, \nonumber\\
h^{2}_{1;\frac{1}{2},-\frac{1}{2}}(J=1)&= \sqrt{2q^2}v, 
\label{lepton helicity}
\end{eqnarray}
where $v=\sqrt{1-4m_{\ell}^2/q^2}$ is the velocity of the lepton in the 
($\ell^+\ell^-$) CM frame, $m_{\ell}$ being the lepton mass. \\ 
As the leptonic current is either purely vector or axial-vector in nature, they 
have definite parity properties which are given by, 
\begin{eqnarray}
h^{1}_{-\lambda_{j};-\lambda_{1},-\lambda_{2}}&=h^{1}_{\lambda_{j};\lambda_{1},\lambda_{2}},\\
h^{2}_{-\lambda_{j};-\lambda_{1},-\lambda_{2}}&=-h^{2}_{\lambda_{j};\lambda_{1},\lambda_{2}}.
\end{eqnarray}

Finally, we move on to the helicity amplitudes  $h_{\lambda_p,0}$ describing the decay $\Lambda \rightarrow p\pi^{-}$. We note that this decay is in itself parity non-conserving in addition to the main decay of $\Lambda_{b}\rightarrow \Lambda+j_{\text{eff}}$. This is in contrast to the well-studied mesonic analouge of $B\rightarrow K^{*}\ell^+ \ell^-$, where the $K^*$ meson subsequently decays to $K\pi$ conserving parity. Also there is only one helicity amplitude for $K^{*}\rightarrow K\pi$ compared to two helicity ampltudes as is the case for the $\Lambda\rightarrow p\pi^-$ decay. The parity non-conserving nature of the $\Lambda \rightarrow p\pi^{-}$ decay leads to the forward-backward asymmetry on the hadron side (angular asymmetry in $\theta_{\Lambda}$) as well as double asymmetries (angular asymmetry in $\theta_{\Lambda}$ and $\theta_{l}$), in addition to the lepton side (angular asymmetry in $\theta_{l}$ ).

\subsection{Full angular distribution}
\label{sec:decay:obs}

\begin{equation}
\begin{aligned}
    \label{eq:angular-distribution}
   &K(q^2,\theta_l,\theta_\Lambda,\phi) \,\cr
     &= \big( K_{1ss} \sin^2\theta_\ell +\, K_{1cc} \cos^2\theta_\ell + K_{1c} \cos\theta_\ell\big) \,\cr
    &  + \big( K_{2ss} \sin^2\theta_\ell +\, K_{2cc} \cos^2\theta_\ell + K_{2c} \cos\theta_\ell\big) \cos\theta_\Lambda
    \cr
    &  + \big( K_{3sc}\sin\theta_\ell \cos\theta_\ell + K_{3s} \sin\theta_\ell\big) \sin\theta_\Lambda \sin\phi\cr
    &  + \big( K_{4sc}\sin\theta_\ell \cos\theta_\ell + K_{4s} \sin\theta_\ell\big) \sin\theta_\Lambda \cos\phi \,.
\end{aligned}
\end{equation}
$K_{1ss}\cdots K_{4s}$ are the angular observables and they are functions 
of $q^2$ and $m_{\ell}$~\cite{Boer:2014kda, Gratrex:2015hna}. We cast this angular distribution in terms of 
orthogonal Legendre functions which is advantageous as the angular observables are uncorrelated to each other. We then provide a set of relations between the 
 $K_{1ss}\cdots K_{4s}$ and the new uncorrelated angular observables $I_{1}\cdots 
I_{10}$ given below,
\begin{eqnarray}
K_{1ss}=I_{1}-\frac{I_{2}}{2}, \ K_{1cc}=I_{1}+I_{2},\nonumber \\
K_{2ss}=I_{4}-\frac{I_{5}}{2},\ K_{2cc}= I_{4}+I_{5},\\
K_{1c}=I_{3},\ K_{2c}=I_{6},\ K_{4sc}=I_{7}, \nonumber \\ 
K_{4s}=I_{8},\ K_{3sc}=I_{9},\ K_{3s}=I_{10}\nonumber.
\end{eqnarray}
The expressions for  $I_{1}\cdots I_{10}$ are derived in terms of the 
transversality amplitudes in Sec.~\ref{sec:helicity amplitudes} and are 
presented in Table~\ref{Table:2}.

In Eq.~\eqref{eq:angular-distribution} full angular analysis is presented from which the complete set of $q^2$ dependent observables are extracted. Once a good 
deal of statistics is available in future it is expected that full reconstruction of the angular observables is possible . For the sake of completeness here we also provide angular observables made of partially 
integrated distributions. Starting from full angular distribution 
(Eq.~\eqref{eq:angular-distribution}) the three uniangular distributions can be 
obtained:

\begin{equation}
\frac{d^2 \Gamma}{d q^2 d\phi} =\frac{1}{4}(16 I_1 + \pi^2 I_8 \cos\phi +\pi^2 I_{10}\sin \phi)
\end{equation}
\begin{equation}
\frac{d^2 \Gamma}{d q^2  \,d \cos\theta_\ell}=-\pi(4I_1 +I_2(1+3\cos 2\theta_\ell) +4 I_3 \cos\theta_\ell  )
\end{equation}
\begin{equation}
\frac{d^2 \Gamma}{d q^2 d \cos\theta_\Lambda}=-4\pi(I_1+I_4\cos\theta_\Lambda)
\end{equation}

\section{ Hadronic helicity amplitudes} 
\label{sec:helicity amplitudes} 

In Eq.\eqref{eq:angular-distribution} we have obtained the angular distribution
of $\Lambda_{b}(\frac{1}{2})\rightarrow\Lambda(\frac{1}{2})+J(0,1)$, where the
$\Lambda$ further decays to $p\pi^-$. Before calculating the helicity amplitudes
of the primary decay let us go through the details of the subsequent hadronic
decay briefly. An onshell spin-$\frac{1}{2}$ $\Lambda$-baryon ($uds$) goes  into
an onshell proton $p$ ($uud$) and a pion $\pi^{-}$ via a parity non-conserving
weak decay that involvles two hadronic couplings $a$ and $b$. The matrix element
for this decay can be written in the following way,
\begin{multline}
 \langle p(k_1) 
\pi(k_2)\vert(\bar{d}\gamma^{\mu}P_{L}u)(\bar{u}\gamma_{\mu}P_{L}s)\vert 
\Lambda(k)\rangle  \\=
\bar{u}(k_1)\Big[ 
\big(a+b\gamma^{5}\big)\Big] u(k).
\end{multline}

We also note that the helicity amplitudes $h_{\lambda_{p},0}$ defined in 
Eq.~\eqref{eq:ang dist first} describe the same decay $\Lambda\rightarrow 
p^+\pi^-$. Moreover it 
is clear that there are only two helicity amplitudes as $\lambda_{p}$ takes 
values $\pm\frac{1}{2}$. From a separate measurement of the $ 
\Lambda\rightarrow p^+\pi^-$ decay width and the polarization asymmetry these 
two helicity amplitudes can be inferred which is equivalent to the extraction 
of the two hadronic couplings $a$ and $b$.

To calculate the hadronic helicity amplitudes of the primary decay which in 
turn can be related to the invariant form factors, we start with the Hamiltonian 
for the decay described in Ref~\cite{Buchalla:1995vs} .

The matrix element for the decay  $\Lambda_b\to \Lambda \bar \ell \ell$ is 
defined by, 
\begin{align}\label{eq:hamiltonian}
\mathcal{M}(\Lambda_b\to &\Lambda \bar \ell \ell)  =  
\frac{G_F}{\sqrt{2}} \frac{ \alpha\lambda_t}{2\,\pi}  
[C_9^{\rm eff}\,
\langle\Lambda\,|\,\bar{s}\,\gamma^\mu (1 - \gamma^5) b\,|\,\Lambda_b\rangle 
\,\bar\ell\gamma_\mu \ell
\nonumber\\
&+
C_{10}\, \langle\Lambda\,|\,\bar{s}\,\gamma^\mu (1 - \gamma^5) 
b\,|\,\Lambda_b\rangle
\, \bar\ell\gamma_\mu \gamma_5 \ell
\\
&- 
\frac{2m_b}{q^2}\,C_7^{\rm eff}\, 
\langle\Lambda\,|\,  \bar{s}\,i\sigma^{\mu 
q}\,(1+\gamma^5)\,\,b\,|\,\Lambda_b\rangle
\,  \bar\ell\gamma_\mu \ell 
] \nonumber\, 
\end{align}
where  $C_{i}'$s are the Wilson coefficients, $\lambda_t\equiv V^{*}_{ts} V_{tb}$ and $m_b$ is the $b$-quark mass. For this paper all the values of the Wilson coefficients have been been taken from Ref ~\cite{Gutsche:2013pp}. The hadronic matrix 
elements written in terms of dimension less form factors as:
\begin{widetext}
\begin{eqnarray}\label{eq:form-factors}
\langle \Lambda(k)\,|\,\bar s\, \gamma^\mu\, b\,| \Lambda_b(p)
\rangle &=&
\bar u_2(p_2)
\Big[ f^V_1(q^2) \gamma^\mu - f^V_2(q^2) i\sigma^{\mu q}/m_{\Lambda_b}
     + f^V_3(q^2) q^\mu/m_{\Lambda_b} \Big] u_1(p_1)\,,
\nonumber\\
\langle  \Lambda(k)\,|\,\bar s\, \gamma^\mu\gamma^5\, b\,| \Lambda_b(p)
\rangle &=&
\bar u_2(p_2)
\Big[ f^A_1(q^2) \gamma^\mu - f^A_2(q^2) i\sigma^{\mu q}/m_{\Lambda_b}
     + f^A_3(q^2) q^\mu/m_{\Lambda_b} \Big]\gamma^5 u_1(p_1)\,,
\nonumber\\
\langle  \Lambda(k)\,|\,\bar s\, i\sigma^{\mu q}/m_{\Lambda_b}\, b\,| 
\Lambda_b(p)
\rangle &=&
\bar u_2(p_2)
\Big[ f^{TV}_1(q^2) (\gamma^\mu q^2 - q^\mu \not\! q)/m_{\Lambda_b}^2
- f^{TV}_2(q^2) i\sigma^{\mu q}/m_{\Lambda_b}\Big] u_1(p_1)\,,
\nonumber\\
\langle  \Lambda(k)\,|\,\bar s\, i\sigma^{\mu q}\gamma^5/m_{\Lambda_b}\, b\,| 
\Lambda_b(p)
\rangle &=&
\bar u_2(p_2)
\Big[ f^{TA}_1(q^2) (\gamma^\mu q^2 - q^\mu \not\! q)/m_{\Lambda_b}^2
- f^{TA}_2(q^2) i\sigma^{\mu q}/m_{\Lambda_b}\Big]\gamma^5 u_1(p_1)\,.
\end{eqnarray}
\end{widetext}
%

\par 
 The helicity amplitudes $H^{a}_{\lambda_2,\lambda_j}$, are expressed by 
 following relation, 
 \begin{equation}
 H^{a}_{\lambda_{\Lambda},\lambda_{j}}= 
 M^{a}_{\mu}(\lambda_{\Lambda})\epsilon^{*\mu}(\lambda_{j}).
 \end{equation}
$M^{a}_{\mu}$ are the hadronic matrix elements defined in
Eq.~\eqref{eq:form-factors}. As before, the labels $\lambda_{j}$ and
$\lambda_{\Lambda}$ denote the helicities of the effective current and daughter
baryon respectively. We shall work in the rest frame of the parent baryon
$\Lambda_{b}$ where the daughter baryon $\Lambda$ moving in the positive $z$
direction and the effective current moving along the negative $z$-axis. The
relevant momenta that describe the motion of particles in this frame are
given below,
 \begin{equation}
p^{\mu}=(m_{\Lambda_b},0,0,0),\ k^{\mu}=(E_2,0,0,p_2), \
q^{\mu}=(q_{0},0,0,-p_2), \nonumber
\end{equation}
where $q_{0}=\frac{1}{2m_{\Lambda_b}}(m_{\Lambda_b}^2-m_{\Lambda}^2+q^2)$ and 
$E_{2}=(m_{\Lambda_b}-q_{0})=(m_{\Lambda_b}^2+m_{\Lambda}^2-q^2)/2m_{\Lambda_b}$.
The helicity of the particles is fixed by angular momentum relation through 
the equation $M_{i}=\lambda_{\Lambda}-\lambda_{j}$.
The $J=\frac{1}{2}$ baryon helicity spinors are given by,
\begin{align}
\bar{u}_{2}\Big(\vec{k}=p_2 \ \hat{z},\pm \frac{1}{2}\Big)&= 
\sqrt{E_2+m_{\Lambda}}\begin{pmatrix}\chi^{\dagger}_{\pm} \frac{\mp\vert p_2 
\vert}{E_2+m_{\Lambda}}\chi^{\dagger}_{\pm}  \end{pmatrix},\\
u_{1}\Big(\vec{p}=0,\pm \frac{1}{2}\Big)&= 
\sqrt{2m_{\Lambda_{b}}}\begin{pmatrix}\chi_{\pm} \\ 0 
\end{pmatrix},
\end{align}
%
where $\chi_{+}=\begin{pmatrix}
1 \\ 0
\end{pmatrix}$ and $\chi_{-}=\begin{pmatrix}
0 \\ 1
\end{pmatrix}$ are two-component Pauli spinors. 
\par
The polarization vectors of the effective current $J_{\text{eff}}$ moving along 
negative $z$-axis look like, 
\begin{equation*}
\epsilon^{\mu}(t)=\frac{1}{\sqrt{q^2}}(q_0,0,0,-p2),
\end{equation*}
\begin{equation}
\epsilon^{\mu}(\pm)=\frac{1}{\sqrt{2}}(0,\pm 1,-i,0)
\end{equation}
\begin{equation*}
\epsilon^{\mu}(0)=\frac{1}{\sqrt{q^2}}(p_2,0,0,-q_{0}).
\end{equation*}
They satisfy the $q_{\mu}\epsilon^{\mu}=0$ equation, $q_{\mu}$ being the 
momentum four-vector of the effective current. 
 We also note that hadronic helicity can be expressed as, 
\begin{equation}
H^{a}_{\lambda_{\Lambda},\lambda_{j}}=H^{Va}_{\lambda_{\Lambda},\lambda_{j}}
-H^{Aa}_{\lambda_{\Lambda},\lambda_{j}}
\end{equation}
where, $H^{Va}_{\lambda_{\Lambda},\lambda_j}$, $H^{Aa}_{\lambda_{\Lambda},\lambda_j}$ are the vector and axial-vector part of the helicity amplitudes respectively. $H^{Va}_{\lambda_{\Lambda},\lambda_j}$, $H^{Aa}_{\lambda_{\Lambda},\lambda_j}$ have definite parity 
properties;
 \begin{equation}
H^{Va}_{-\lambda_{\Lambda},-\lambda_j}=H^{Va}_{\lambda_{\Lambda},\lambda_j} 
\qquad 
H^{Aa}_{-\lambda_{\Lambda},-\lambda_j}=-H^{Aa}_{\lambda_{\Lambda},\lambda_j}.
\end{equation}
Different hadronic helicity amplitudes that take part in the decay are presented below,
\begin{eqnarray}
\label{eq:H-F-relation}
H^{Va}_{\frac{1}{2} t} &=& 
\sqrt{\frac{Q_+}{q^2}} \, 
\biggl( M_- \, F_1^{Va} + \frac{q^2}{m_{\Lambda_b}} \, F_3^{Va} \biggr)\,, 
\nonumber\\
H^{Va}_{\frac{1}{2} 1} &=& \sqrt{2 Q_-} \, 
\biggl( F_1^{Va} + \frac{M_+}{m_{\Lambda_b}} \, F_2^{Va} \biggr)\,, \nonumber\\
H^{Va}_{\frac{1}{2} 0} &=& \sqrt{\frac{Q_-}{q^2}} \,  
\biggl( M_+ \, F_1^{Va} + \frac{q^2}{m_{\Lambda_b}} \, F_2^{Va} \biggr)\,, 
\\
H^{Aa}_{\frac{1}{2} t} &=& \sqrt{\frac{Q_-}{q^2}} \, 
\biggl( M_+ \, F_1^{Aa} - \frac{q^2}{m_{\Lambda_b}} \, F_3^{Aa} \biggr)\,, 
\nonumber\\
H^{Aa}_{\frac{1}{2} 1} &=& \sqrt{2 Q_+} \, 
\biggl( F_1^{Aa} - \frac{M_-}{m_{\Lambda_b}} \, F_2^{Aa} \biggr)\,, \nonumber\\
H^{Aa}_{\frac{1}{2} 0} &=& \sqrt{\frac{Q_+}{q^2}} \,  
\biggl( M_- \, F_1^{Aa}  - \frac{q^2}{m_{\Lambda_b}} \, F_2^{Aa} \biggr)\,, 
\nonumber  
\end{eqnarray}
where $M_\pm = m_{\Lambda_b} \pm m_{\Lambda}$, $Q_\pm = M_\pm^2 - q^2$ and $a$ 
being the leptonic current index ($a=1$; vector current, $a=2$; axial-vector 
current).  
 The  redefined form factors $F^{Va}_i$, $F^{Aa}_i$ involve linear combinations 
of 
the form factors $f_i^V$, $f_i^A$  as well as the Wilson 
coefficients.
\begin{table*}[thb]
\centering
\resizebox{\textwidth}{!}{%
\begin{tabular}{c c c }
\hline
\hline
 Label&  Angular Term&  Transversity amplitude   \\
 \hline
$I_{1}$ & Cosnt. & $\tau q^2\Big[\frac{2}{3}\big(1-\frac{m_{\ell}^2}{q^2}\big) 
\big\{ \vert A^{L}_{\parallel,0}\vert^{2}+\vert A^{L}_{\parallel,1} 
\vert^{2}+\vert A^{L}_{\perp,0} \vert^{2}+\vert A^{L}_{\perp,1} \vert^{2}+ 
\big( L \leftrightarrow R \big)\big\}$ \\ & & + 
$\frac{4m_{\ell}^2}{q^2}\text{Re}\big(A^{*R}_{\parallel,0}A^{L}_{\parallel,0}+A^{*R}_{\parallel,1}A^{L}_{\parallel,1}+A^{*R}_{\perp,0}A^{L}_{\perp,0}+A^{*R}_{\perp,1}A^{L}_{\perp,1}
 \big)$ \\ & & $+\frac{2m_{\ell}^2}{q^2}\big(\vert A_{t,\perp}\vert^2+\vert 
 A_{t,\parallel}\vert^2 \big)\Big]  $ \\

 $I_2$ & $P_{2}(\cos \theta_l)=\frac{1}{2}(3 \cos \theta_{l}^2-1)$ & $ 
 \frac{q^2 \tau}{3}\big(1-\frac{4m_{\ell}^2}{q^2}\big)\Big[\vert 
 A^{L}_{\parallel,1} \vert^{2}+\vert A^{L}_{\perp,1}\vert^{2}-2(\vert 
 A^{L}_{\parallel,0} \vert^{2}+\vert A^{L}_{\perp,0} \vert^{2})+\big( L 
 \leftrightarrow R \big) \Big]$ \\

$I_3$& $P_{1}(\cos \theta_l)= \cos \theta_{l}$ & $-2 q^2 \tau v \ 
\text{Re}\Big[ A^{*L}_{\perp,1}A^{L}_{\parallel,1}-\big( L \leftrightarrow R 
\big)\Big]$ \\

$I_4$& $P_{1}(\cos \theta_{\Lambda})= \cos \theta_{\Lambda}$ & $\frac{4}{3}q^2 
\beta 
\Big[(1-\frac{m_{\ell}^2}{q^2})\text{Re}\Big\{A^{*L}_{\perp,0}A^{L}_{\parallel,0}+A^{*L}_{\perp,1}A^{L}_{\parallel,1}+
 \big( L \leftrightarrow R \big)\Big\}$ \\ & & $+\frac{3m_{\ell}^2}{q^2}\big\{ 
\text{Re} 
(A^{*R}_{\perp,0}A^{L}_{\parallel,0}+A^{*L}_{\perp,0}A^{R}_{\parallel,0}+A^{*R}_{\perp,1}A^{L}_{\parallel,1}+A^{*L}_{\perp,1}A^{R}_{\parallel,1})
 \big\} $ \\ & & $+ \frac{3m_{\ell}^2}{q^2}\text{Re}\big[ A^{*}_{t,\parallel} 
 A_{t,\perp}\big] \Big]$ \\ 

$I_5$& $P_{2}(\cos \theta_l)P_{1}(\cos \theta_{\Lambda})$ & \\ & 
=$\frac{1}{2}(3 \cos \theta_{l}^2-1)\cos \theta_{\Lambda}$ & $\frac{2}{3} q^2 
\beta(1-\frac{4m_{\ell}^2}{q^2})\text{Re}\Big[A^{*L}_{\perp,1}A^{L}_{\parallel,1}-2A^{*L}_{\perp,0}A^{L}_{\parallel,0}+\big(
 L \leftrightarrow R \big) \Big]$\\

$I_6$& $P_{1}(\cos \theta_l)P_{1}(\cos \theta_{\Lambda})$ & \\ & =$\cos 
\theta_{l}\cos \theta_{\Lambda}$ & $ -q^2 v \beta \Big[ \vert 
A^{L}_{\parallel,1} \vert^{2}+\vert A^{L}_{\perp,1} \vert^{2}-\big( L 
\leftrightarrow R \big) \Big]$\\

$I_7$& $P_{1}(\cos \theta_l)\sin \theta_{l}\sin \theta_{\Lambda} \cos \phi$ & 
\\ & =$\cos \theta_{l}\sin \theta_{l}\sin \theta_{\Lambda}\cos \phi$ & $ 
\sqrt{2} q^2\beta(1-\frac{4m_{\ell}^2}{q^2})\ \text{Re} 
\Big[\big(A_{\perp,1}^{*L}A_{\parallel,0}^{L} 
-A_{\parallel,1}^{*L}A_{\perp,0}^{L}  \big) +\big( L \leftrightarrow R \big) 
\Big]$\\

$I_8$& $\sin \theta_{l}\sin \theta_{\Lambda} \cos \phi$ & $ \sqrt{2} q^2v\beta 
\  \text{Re} \Big[\big(A_{\perp,1}^{*L}A_{\perp,0}^{L} 
-A_{\parallel,1}^{*L}A_{\parallel,0}^{L}  \big) -\big( L \leftrightarrow R 
\big) \Big]$\\

$I_9$& $P_{1}(\cos \theta_l)\sin \theta_{l}\sin \theta_{\Lambda} \sin \phi$ & 
\\ & =$\cos \theta_{l}\sin \theta_{l}\sin \theta_{\Lambda}\sin \phi$ & $- 
\sqrt{2} q^2\beta(1-\frac{4m_{\ell}^2}{q^2}) \ \text{Im} 
\Big[\big(A_{\perp,1}^{*L}A_{\perp,0}^{L} 
-A_{\parallel,1}^{*L}A_{\parallel,0}^{L}  \big) +\big( L \leftrightarrow R 
\big) \Big]$\\

$I_{10}$& $\sin \theta_{l}\sin \theta_{\Lambda} \sin \phi$    & $- \sqrt{2} 
q^2v\beta \ \text{Im} \Big[\big(A_{\perp,1}^{*L}A_{\parallel,0}^{L} 
-A_{\parallel,1}^{*L}A_{\perp,0}^{L}  \big) -\big( L \leftrightarrow R \big) 
\Big]$\\
\hline
\hline
 \end{tabular}
}
\caption{Angular observables expressed in terms of transversity amplitudes 
defined in Sec.~\ref{sec:helicity amplitudes} (see Eqs. 
~\eqref{eq:Aa0}--\eqref{eq:Apt}). 
$\tau$ and $\beta$ are the 
total decay rate and the forward-backward asymmetry of the subsequent hadronic 
decay of $\Lambda$ to $p\pi$ respectively.}
\label{Table:2}
\end{table*}


\begin{eqnarray}\label{eq:wilson-coeff}
F_1^{V1} &=& 
C_9^{\rm eff} \, f_1^V 
       - \frac{2 m_b}{m_{\Lambda_b}} \, C_7^{\rm eff} \, f_1^{TV} 
\,, 
\nonumber\\
F_2^{V1} &=& 
C_9^{\rm eff} \, f_2^V 
       - \frac{2 m_b m_{\Lambda_b}}{q^2} \, C_7^{\rm eff} \, f_2^{TV} 
\,, 
\nonumber\\
F_3^{V1} &=& 
C_9^{\rm eff} \, f_3^V 
       + \frac{2 m_b M_-}{q^2} \, C_7^{\rm eff} \, f_1^{TV} 
\,, 
\nonumber\\
F_1^{A1} &=& 
C_9^{\rm eff} \, f_1^A 
      + \frac{2 m_b}{m_{\Lambda_b}} \, C_7^{\rm eff} \, f_1^{TA} 
\,, 
\nonumber\\
F_2^{A1} &=& 
C_9^{\rm eff} \, f_2^A 
      + \frac{2 m_b m_{\Lambda_b}}{q^2} \, C_7^{\rm eff} \, f_2^{TA} 
\,, 
\nonumber\\
F_3^{A1} &=& 
C_9^{\rm eff} \, f_3^A 
       + \frac{2 m_b M_+}{q^2} \, C_7^{\rm eff} \, f_1^{TA} 
\,, \nonumber 
\end{eqnarray}
and 
\begin{eqnarray}
F_i^{V2} &=& C_{10} \, f_i^{V}\,, \nonumber\\
F_i^{A2} &=& C_{10} \, f_i^{A}\,.  
\end{eqnarray}
We switch to transversity amplitue defined as:
\begin{eqnarray}
\label{eq:Aa0}
A_{\parallel,0}^{L(R)} &=& H^{V a=1}_{\frac{1}{2},0}   \mp  H^{V 
a=2}_{\frac{1}{2},0}\\ 
\label{eq:Ap0} \nonumber \\
A_{\perp,0}^{L(R)} &=& H^{A a=1}_{\frac{1}{2},0} \mp H^{A 
a=2}_{\frac{1}{2},0}\\ \label{eq:Aa1} \nonumber \\
A_{\parallel,1}^{L(R)} &=& H^{V a=1}_{\frac{1}{2},1}  \mp H^{V 
a=2}_{\frac{1}{2},1}\\ \label{eq:Ap1} \nonumber \\
A_{\perp,1}^{L(R)} &=& H^{A a=1}_{\frac{1}{2},1} \mp H^{A a=2}_{\frac{1}{2},1}
\\\label{eq:Aat} \nonumber \\
A_{\parallel,t}&=& H^{a=2}_{-\frac{1}{2},t}+H^{a=2}_{\frac{1}{2},t},  \\
\label{eq:Apt} \nonumber \\
A_{\perp,t}&=& H^{a=2}_{-\frac{1}{2},t}-H^{a=2}_{\frac{1}{2},t}, 
\end{eqnarray}

The superscript $L$($R$) on $A_{\perp(\parallel)}$ denotes that the transversity
amplitudes are multiplied by left-handed (right-handed) lepton current. There
are two additional transversity amplitudes that are relevant to the decay if the
$j_{\text{eff}}$ is virtual, corresponding to the $J=0$ contribution. These two
amplitudes $A_{\parallel,t}$ and $A_{\perp,t}$ do not have separate left-handed
or right-handed part as the timelike polarization of $j_{\text{eff}}$ couples
only to the axial-vector part of the lepton current~\cite{Altmannshofer:2008dz}, a fact highlighted by
Eq.~\eqref{lepton helicity}. Moreover, the $A_{t}$ contribution vanishes in the
limit of massless leptons. The decay $\Lambda_b \to \Lambda (\to p \pi)
\ell^+\ell^-$ is completely described by these transversity amplitudes which
include all contribution from the standard model effective operators. 

\begin{table*}[bht]
\begin{tabular}{c l l l l l l l l l l}
\hline
\hline
Parameter & $f_1^V$ & $f_2^V$ & $f_3^V$ 
& $f_1^A$ & $f_2^A$ & $f_3^A$
& $f_1^{TV}$ & $f_2^{TV}$ 
& $f_1^{TA}$ & $f_2^{TA}$ 
\\
\hline
$f(0)$ & 0.107 & 0.043 & 0.003 & 0.104 & 0.003 & -0.052 & -0.043 & -0.105 & 
0.003 & -0.105\\

$a$    & 2.271 & 2.411 & 2.815 & 2.232 & 2.955 & 2.437  & 2.411  & 0.072  & 
2.955  & 2.233\\

$b$    & 1.367 & 1.531 & 2.041 & 1.328 & 3.620 & 1.559  & 1.531  & 0.001  & 
3.620  & 1.328\\
\hline
\hline
\end{tabular}
\caption{Parameters for the form factors as a function of $q^2$, 
$f(t)=f(0)/(1-at+bt^2)$, $t=q^2/m_{\Lambda_b}^2$ for $\Lambda_{b}\rightarrow 
\Lambda$ transition as given in Ref.~\cite{Gutsche:2013pp}}
\label{Table:3}
\end{table*}

\section{Total decay rate and angular observables}
\label{Decay rate and angular observables}

The total differential decay rate can be extracted in terms of the constant 
piece appearing in the angular distribution once we include the parameters in 
the effective Hamiltonian and the relevant phase space factors, i.e. 
\begin{widetext}
\begin{align}
\frac{d\Gamma}{dq^2}&\equiv\frac{d\Gamma(\Lambda_b \rightarrow (\Lambda\to 
p\pi)\ell^+ 
\ell^-)}{dq^2} \nonumber\\  
 &\dsp=
 \text{Br}(\Lambda\rightarrow p \pi^-) 
 \times
\frac{1}{2}\frac{1}{(2\pi)^3}\frac{\vert\textbf{p}_{\textbf{2}} \vert q^2 
v}{16m_{\Lambda_b}^2} \Big(\frac{G_F \alpha\lambda_t}{2\,\pi}\Big)^2 \times 
\Big[\frac{2}{3}\dsp\big(1-\frac{m_{\ell}^2}{q^2}\big) 
\big\{ \vert A^{L}_{\parallel,0}\vert^{2} 
+\vert A^{L}_{\parallel,1} 
\vert^{2}+\vert A^{L}_{\perp,0} \vert^{2}+\vert A^{L}_{\perp,1} \vert^{2} + 
\big( L \leftrightarrow R \big)\big\} \nonumber \\
&\qquad +
\frac{4m_{\ell}^2}{q^2}\text{Re}\big(A^{*R}_{\parallel,0}A^{L}_{\parallel,0}
+A^{*R}_{\parallel,1}A^{L}_{\parallel,1}+A^{*R}_{\perp,0}A^{L}_{\perp,0} 
+A^{*R}_{\perp,1}A^{L}_{\perp,1}
 \big)+\frac{2m_{\ell}^2}{q^2}\big(\vert A_{t,\perp}\vert^2+\vert 
 A_{t,\parallel}\vert^2 \big)\Big],
 \end{align}
 \end{widetext}
where $\alpha$ is the fine structure constant, $G_{F}$ is the Fermi coupling constant, 
$\lambda_t=V_{ts}^{\dagger}V_{tb}$ is the product of CKM matrix elements 
relevant for the underlying quark level transition and $\vert 
\textbf{p}_{\textbf{2}}\vert=\lambda^{1/2}(m_{\Lambda_b}^2,m_{\Lambda}^2,q^2)/2 
m_{\Lambda_b}$ is the momentum of $\Lambda$-baryon in the $\Lambda_{b}$ rest 
frame where $\lambda^{1/2}(m_{\Lambda_b}^2,m_{\Lambda}^2,q^2)$ is the K\"all\'en Function. The 1/2 factor appearing in the definition of the differential decay rate 
takes into account the decay of unpolarized spin-1/2 initial state 
$\Lambda_{b}$-baryon. We note that there is an additional timelike contribution 
to the differential decay rate that becomes important for non-zero lepton 
masses $m_{\ell} \neq 0$ especially in the low-$q^2$ region. 
\begin{figure}[b!]
\centering 
\includegraphics[width=0.45\textwidth]{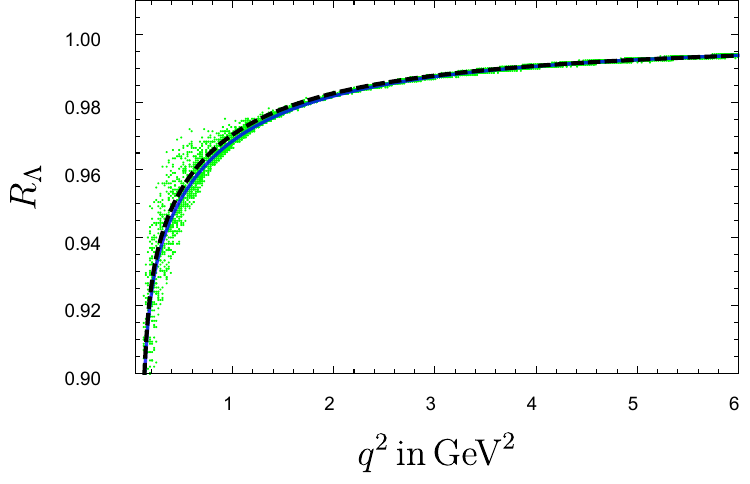} 
\caption{The $q^2$-dependence of $R_{\Lambda}$. The blue solid line is found using
the values of the form factors in covariant quark model given in
Table~\ref{Table:3}. This line almost coincides with the estimate using SCET form
factors plotted as a black dotted line. The green band represents the possible 
values of $R_\Lambda$ obtained by
randomly generating  $10^4$ points corresponding to $\pm 30\%$ error in
each of the covariant
quark model form factor estimates.} 
\label{Fig:1} 
\end{figure}

The decay rates for $\Lambda_{b}\to \Lambda e^+e^-$ and $\Lambda_{b}\to \Lambda
\mu^+\mu^-$ can be readily calculated once the values of relevant form factors
are fixed. Our interest lies primarily in low-$q^2$ (high recoil) region i.e.
$q^2=0.04\gev^2-6\gev^2$ and it has been emphasized in ~\cite{Korner:1992uw,
Korner:1991zx, Isgur:1990jg, Mannel:2011xg, Dosch:1997zx} that heavy-quark
symmetry is not reliable at low-$q^2$. Heavy quark symmetry is expected to break
down as one deviates from the zero-recoil point ~\cite{Korner:1992uw,
Korner:1991zx}. We, therefore, follow the approach of~\cite{Gutsche:2013pp},
which uses covariant quark model (CQM)  to calculate the required form factors
at low-$q^2$. We quote the values of those form factors \cite{Gutsche:2013pp} in
Table~\ref{Table:3} and calculate $R(\Lambda)$, the ratio of decay rates for
$\Lambda_{b}\to \Lambda \mu^+\mu^-$ and $\Lambda_{b}\to \Lambda e^+e^-$. We 
also, use Soft-collinear effective
theory (SCET) and compare the two estimates obtained for the decay rate.
A heavy-to-light transition of $\Lambda_{b}$ to $\Lambda$ in large recoil (low
$q^2$) limit is simplified as the number of independent form factor reduces to
one. SCET  is valid ~\cite{Wang:2011uv, Mannel:2011xg} in this energy range as
the energy of the daughter $\Lambda$ is larger than its mass and one can use the
$\lambda=\sqrt{\frac{m_{\Lambda}}{m_{b}}}$ as an expansion parameter which is
small. In such a picture, the hadronic matrix elements in
Eq.~\eqref{eq:form-factors} for $\Lambda_{b} \to \Lambda$ decay can be
parametrized in the following way,
\begin{eqnarray}
\langle \Lambda(p_2) \vert \bar{s}\Gamma b \vert \Lambda_{b}(p) \rangle \simeq 
\xi_{\lambda}(E_{2})\bar{u}_{\Lambda}(p_2) \Gamma 
u_{\Lambda_{b}}(p)+\mathcal{O}(\lambda^2\xi_{\lambda})\nonumber
\end{eqnarray}
To start with, let us highlight the relations between different form factors 
used in Eqn. \eqref{eq:form-factors} in low $q^2$ limit, \begin{eqnarray}
f_{1}^{V}\approx f_{1}^{A}\approx -f_{2}^{TV}\approx -f_{2}^{TA}=\xi_{\lambda} 
\\
f_{2}^{V}\approx f_{3}^{V}\approx f_{2}^{A} \approx f_{3}^{A}\approx 
f_{1}^{TV}\approx f_{1}^{TA}\approx 0
\end{eqnarray}
where $\xi_{\lambda}$ is the single parameter all non-zero form factors depend
on in the limit of small $q^2$~\cite{Feldmann:2011xf, Wang:2011uv}. In
Fig.~\ref{Fig:1}, we have plotted $R(\Lambda)$. We have gone further and also
probed the reliability of  $R_\Lambda$ value by randomly adding $\pm 30\%$
error to each form factor estimate in covariant quark model and by generating 
$10^4$
points to evaluate the ratio.  It is thus concluded that $R_\Lambda$ is reliably
predicted in the low-$q^2$ (high recoil) region of $q^2=0.04\gev^2-6\gev^2$. 
The contributions to $R_{\Lambda}$ from long-distance effects will be discussed later.

\begin{figure*}[hbt]
\centering 
\includegraphics[width=0.4\textwidth]{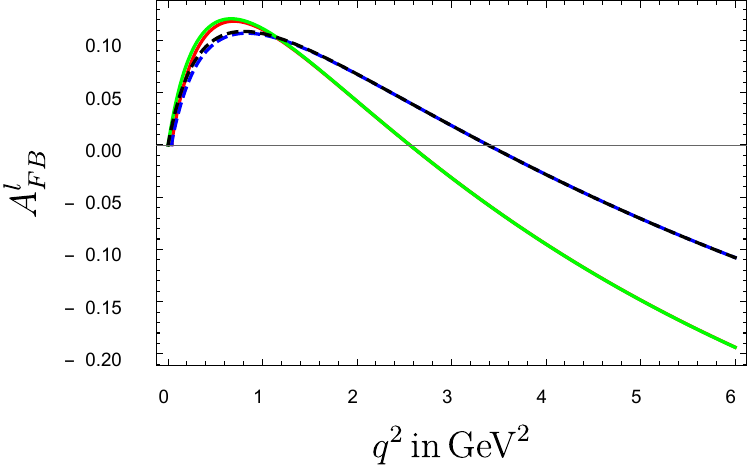}
\includegraphics[width=0.4\textwidth]{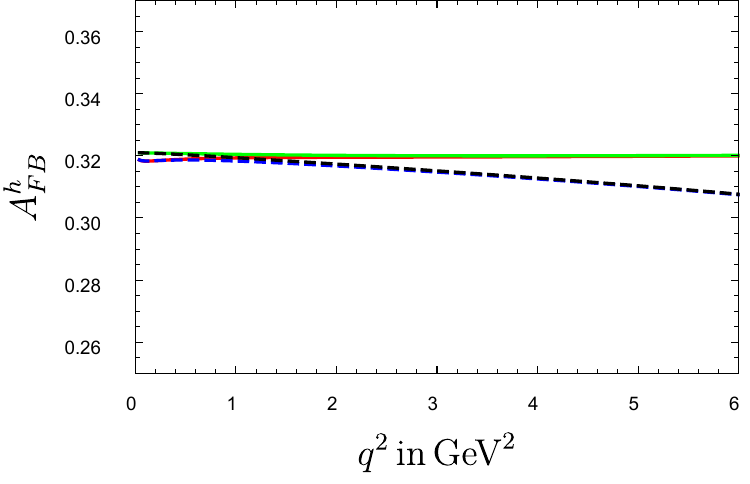}
\includegraphics[width=0.4\textwidth]{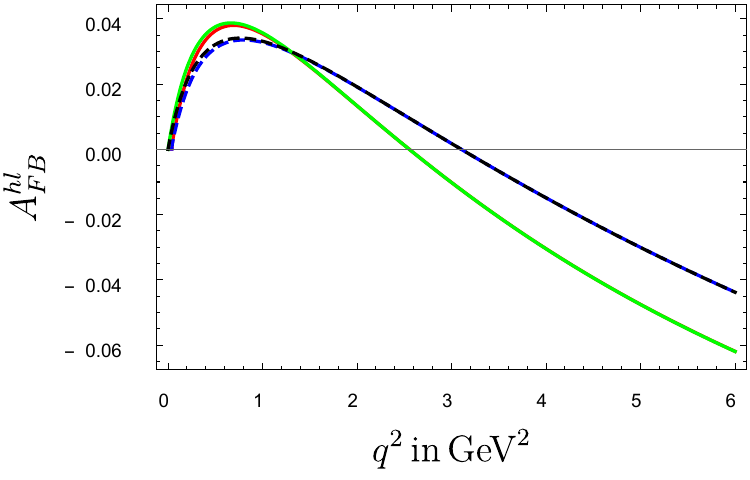}
\caption{The $q^2$-dependence of the forward-backward asymmetries  $A_{FB}^{l}$,
$A_{FB}^{h}$, $A_{FB}^{hl}$ for electron and muon are presented. For muons, the $q^2$-dependence of the asymmetries are given by the solid red line, obtained using covariant quark model form-factors and the blue dotted line is for SCET form factors. For electrons, the solid green line represents the $q^2$-dependence using covariant quark model form factors whereas the black dotted line represents the $q^2$-dependence due to SCET form factors.}
\label{Fig:2}
\end{figure*}

\subsection{Angular Observables}
\label{angular observables}
In this section we list the angular asymmetries that allow us to extract the 
angular coefficients $I_{2} \cdots I_{10}$ and contribute to nine of the ten 
observables, with $I_1$ being the total differential decay rate. These 
asymmetries result from orthogonal angular distribution and are thus 
independent observables.
\begin{widetext}
\begin{align}
\label{eq:asymmetry}
A_2 &=\frac{ \dsp\Big[ 
\int_{-1}^{-\frac{1}{2}}-\int_{-\frac{1}{2}}^{\frac{1}{2}}+\int_{\frac{1}{2}}^{1}
 \Big] d\cos \theta_l \int_{-1}^{1} d\cos \theta_{\Lambda} \int_{0}^{2\pi}d\phi 
\frac{1}{3\pi}\frac{d^4\Gamma}{dq^2 d\cos \theta_{\Lambda}  d\cos \theta_l 
d\phi}}{\dsp\int_{-1}^{1}d\cos \theta_l \int_{-1}^{1} d\cos 
\theta_{\Lambda}\int_{0}^{2\pi}d\phi\frac{d^4\Gamma}{dq^2 d\cos 
\theta_{\Lambda}  d\cos \theta_l d\phi}}\\
A_3 &=\frac{\dsp\Big[-\int_{-1}^{0}+\int_{0}^{1}\Big]d\cos \theta_l 
\int_{-1}^{1} 
d\cos \theta_{\Lambda}\int_{0}^{2\pi}d\phi \frac{1}{4\pi}\frac{d^4\Gamma}{dq^2 d\cos 
\theta_{\Lambda}  d\cos \theta_l d\phi}}{\dsp\int_{-1}^{1}d\cos \theta_l 
\int_{-1}^{1} d\cos \theta_{\Lambda}\int_{0}^{2\pi}d\phi\frac{d^4\Gamma}{dq^2 
d\cos \theta_{\Lambda}  d\cos \theta_l d\phi}}\\
A_4 &=\frac{\dsp\int_{-1}^{1} d\cos \theta_l 
\Big[-\int_{-1}^{0}+\int_{0}^{1}\Big]d\cos \theta_{\Lambda}\int_{0}^{2\pi}d\phi 
\frac{1}{4\pi}\frac{d^4\Gamma}{dq^2 d\cos \theta_{\Lambda}  d\cos \theta_l 
d\phi}}{\dsp\int_{-1}^{1}d\cos \theta_l \int_{-1}^{1} d\cos 
\theta_{\Lambda}\int_{0}^{2\pi}d\phi\frac{d^4\Gamma}{dq^2 d\cos 
\theta_{\Lambda}  d\cos \theta_l d\phi}}\\
A_5 
&=\frac{4}{3}\frac{\dsp\Big[-\int_{-1}^{-\frac{1}{2}}+\int_{-\frac{1}{2}}^{\frac{1}{2}}-\int_{\frac{1}{2}}^{1}
 \Big] d\cos \theta_l \Big[\int_{-1}^{0}-\int_{0}^{1}\Big] d\cos 
\theta_{\Lambda}\int_{0}^{2\pi}d\phi \frac{d^4\Gamma}{dq^2 d\cos \theta_{\Lambda}  d\cos \theta_l 
d\phi}}{\dsp\int_{-1}^{1}d\cos \theta_l \int_{-1}^{1} d\cos 
\theta_{\Lambda}\int_{0}^{2\pi}d\phi\frac{d^4\Gamma}{dq^2 d\cos 
\theta_{\Lambda}  d\cos \theta_l d\phi}}\\
A_6 &= \frac{\dsp\Big[\int_{-1}^{0}-\int_{0}^{1}\Big]d\cos \theta_l 
\Big[\int_{-1}^{0}-\int_{0}^{1}\Big] d\cos 
\theta_{\Lambda}\int_{0}^{2\pi}d\phi \frac{d^4\Gamma}{dq^2 d\cos \theta_{\Lambda}  d\cos \theta_l 
d\phi}}{\dsp\int_{-1}^{1}d\cos \theta_l \int_{-1}^{1} d\cos 
\theta_{\Lambda}\int_{0}^{2\pi}d\phi\frac{d^4\Gamma}{dq^2 d\cos 
\theta_{\Lambda}  d\cos \theta_l d\phi}}\\
A_7 &=-\frac{3}{4}\frac{\dsp\Big[\int_{-1}^{0}-\int_{0}^{1}\Big]d\cos \theta_l 
\int_{-1}^{1}d\cos 
\theta_{\Lambda}\Big[\int_{\frac{-\pi}{2}}^{\frac{\pi}{2}}-\int_{\frac{\pi}{2}}^{\frac{3\pi}{2}}\Big]d\phi\frac{d^4\Gamma}{dq^2
 d\cos \theta_{\Lambda}  d\cos \theta_l d\phi}}{\dsp\int_{-1}^{1}d\cos \theta_l 
\int_{-1}^{1} d\cos \theta_{\Lambda}\int_{0}^{2\pi}d\phi\frac{d^4\Gamma}{dq^2 
d\cos \theta_{\Lambda}  d\cos \theta_l d\phi}}\\
A_8 &=\frac{\dsp\int_{-1}^{1}d\cos \theta_l \int_{-1}^{1}d\cos 
\theta_{\Lambda}\Big[-\int_{-\pi}^{\frac{-\pi}{2}}+\int_{\frac{-\pi}{2}}^{\frac{\pi}{2}}-\int_{\frac{\pi}{2}}^{\pi}\Big]d\phi\frac{d^4\Gamma}{dq^2
 d\cos \theta_{\Lambda}  d\cos \theta_l d\phi}}{\dsp\int_{-1}^{1}d\cos \theta_l 
\int_{-1}^{1} d\cos \theta_{\Lambda}\int_{0}^{2\pi}d\phi\frac{d^4\Gamma}{dq^2 
d\cos \theta_{\Lambda}  d\cos \theta_l d\phi}}\\
A_9 &=-\frac{3}{4}\frac{\Big[\dsp\int_{-1}^{0}-\int_{0}^{1}\Big]d\cos \theta_l 
\int_{-1}^{1}d\cos 
\theta_{\Lambda}\Big[\int_{0}^{\pi}-\int_{\pi}^{2\pi}\Big]d\phi\frac{d^4\Gamma}{dq^2
 d\cos \theta_{\Lambda}  d\cos \theta_l d\phi}}{\dsp\int_{-1}^{1}d\cos \theta_l 
\int_{-1}^{1} d\cos \theta_{\Lambda}\int_{0}^{2\pi}d\phi\frac{d^4\Gamma}{dq^2 
d\cos \theta_{\Lambda}  d\cos \theta_l d\phi}}\\
A_{10} &=\frac{1}{\pi^2}\frac{\dsp\int_{-1}^{1}d\cos \theta_l \int_{-1}^{1}d\cos 
\theta_{\Lambda}\Big[\int_{-\pi}^{0}+\int_{0}^{\pi}\Big]d\phi\frac{d^4\Gamma}{dq^2
 d\cos \theta_{\Lambda}  d\cos \theta_l d\phi}}{\dsp\int_{-1}^{1}d\cos \theta_l 
\int_{-1}^{1} d\cos \theta_{\Lambda}\int_{0}^{2\pi}d\phi\frac{d^4\Gamma}{dq^2 
d\cos \theta_{\Lambda}  d\cos \theta_l d\phi}}\
\end{align}
\end{widetext}

Note that $A_9$ and $A_{10}$ are non-zero only if the amplitudes have imaginary
contributions. These are expected to be extremely tiny in the SM. The 
asymmetries $A_2$, $A_5$, $A_7$ and $A_8$ are not simple forward back 
asymmetries.  We note that
$A_{3}$, $A_4$  are forward-backward asymmetries in the leptonic angle
$\theta_{l}$ and hadronic angle $\theta_{\Lambda}$ respectively. There is also a
double asymmetry involving $\theta_{l}$ and $\theta_{\Lambda}$ given by $A_{6}$.
We provide an expression for each of these quantities in terms of known
parameters. 
\begin{figure*}[t]
\centering 
\includegraphics[width=0.4\textwidth]{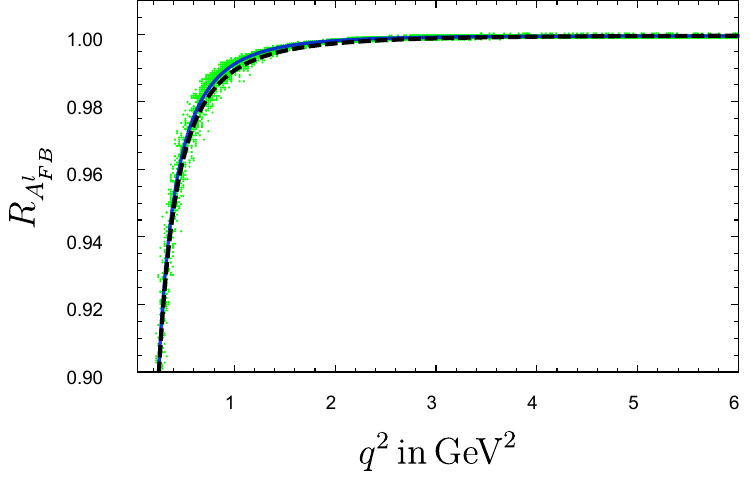}
\includegraphics[width=0.4\textwidth]{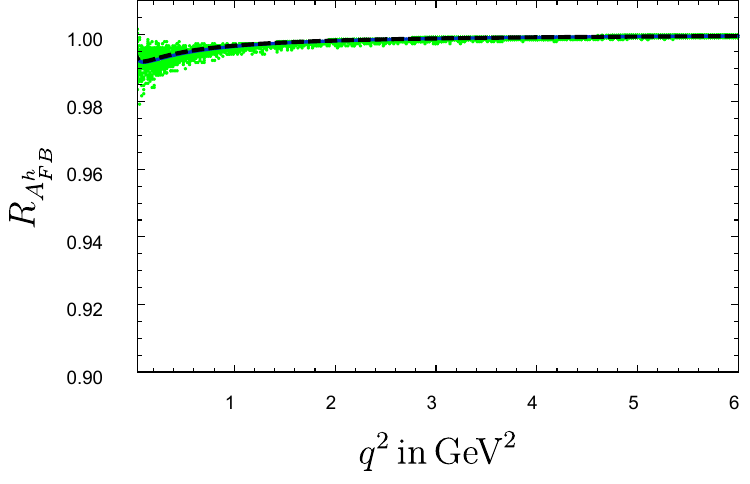}
\includegraphics[width=0.4\textwidth]{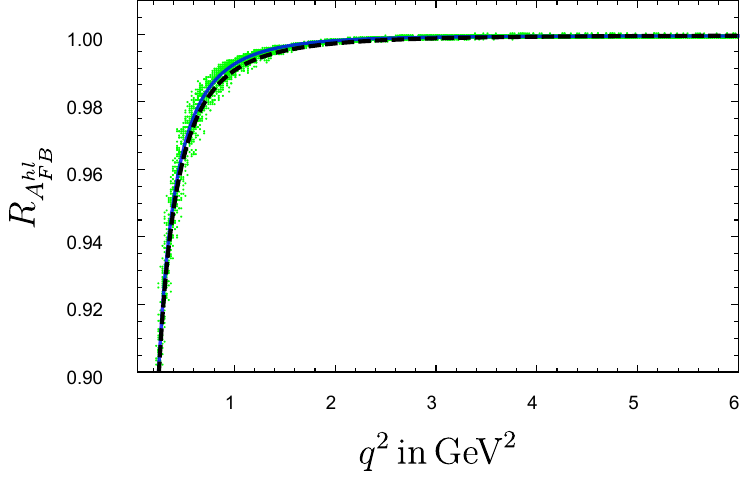}

\caption{The $q^2$ dependence for the ratios of forward-backward asymmetries
$R_{A_{FB}^{l}}$, $R_{A_{FB}^{h}}$ and $R_{A_{FB}^{hl}}$ (see
Eqs.~\eqref{eq:Rl2}--\eqref{eq:Rhl}). The blue solid line is found using the values
of form factors in covariant quark model given in Table~\ref{Table:3}. This
line almost coincides with the estimate using SCET form factors, represented as the black dotted line. Note that, while the
two asymmetries $A_{FB,\mu}^{h}$  and $A_{FB,e}^{h}$ are form factor dependent
the ratio $R_{A_{FB}^{h}}$ is independent of the choice of form-factors (CQM or
SCET). The green band represents the possible 
values of $R_\Lambda$ obtained by
randomly generating  $10^4$ points corresponding to $\pm 30\%$ error in
each of the covariant
quark model form factor estimates. }
\label{Fig:3} 
\end{figure*}  

\begin{widetext}
\begin{eqnarray}
A^{l}_{FB}&=&\dsp\frac{-3v \ 
\text{Re}\big[A^{*L}_{\perp,1}A^{L}_{\parallel,1}-\big( 
L 
\leftrightarrow R 
\big) \big]q^2}{4 I_1}\label{ratio1}\\
\label{ratio2}
A^{h}_{FB}&=& \dsp\frac{\dsp \alpha_{\Lambda} \Bigg[ \splitfrac{
(1-\frac{m_{\ell}^2}{q^2})\text{Re}\Big\{A^{*L}_{\perp,0}A^{L}_{\parallel,0}
+A^{*L}_{\perp,1}A^{L}_{\parallel,1}+
 \big( L \leftrightarrow R \big) \Big \}+ 
\frac{3m_{\ell}^2}{q^2}\text{Re}\{ A^{*}_{t,\parallel} A_{t,\perp}\} 
 }{\dsp+\frac{3m_{\ell}^2}{q^2}\big\{ \text{Re} 
(A^{*R}_{\perp,0}A^{L}_{\parallel,0}+A^{*L}_{\perp,0}A^{R}_{\parallel,0}+A^{*R}_{\perp,1}A^{L}_{\parallel,1}+A^{*L}_{\perp,1}A^{R}_{\parallel,1})
 \big\}}\Bigg]q^2 }{{2I_1}}\\
\label{ratio3}
 A^{hl}_{FB}&=&\dsp\frac{- 3v \alpha_{\Lambda} \Big[ \vert 
 A^{L}_{\parallel,1} \vert^{2}+\vert A^{L}_{\perp,1} \vert^{2}-\big( L 
 \leftrightarrow R \big) \Big]q^2}{8 I_1}
\end{eqnarray}
\end{widetext}
\begin{figure*}[t]
\includegraphics[width=0.4\textwidth]{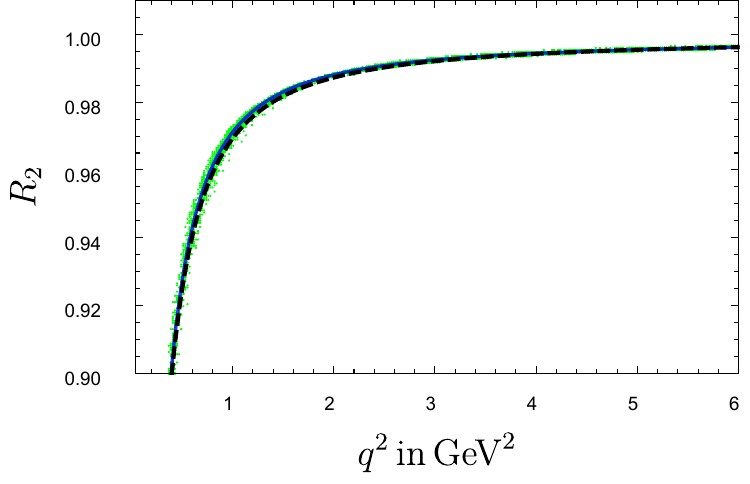}
\includegraphics[width=0.4\textwidth]{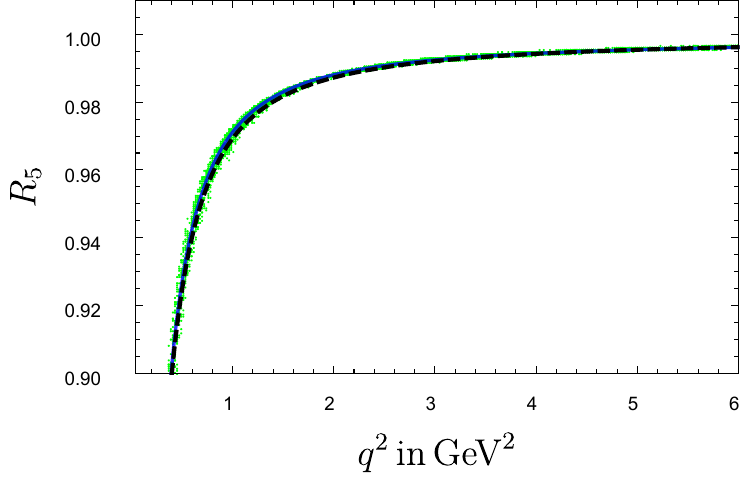}
\includegraphics[width=0.4\textwidth]{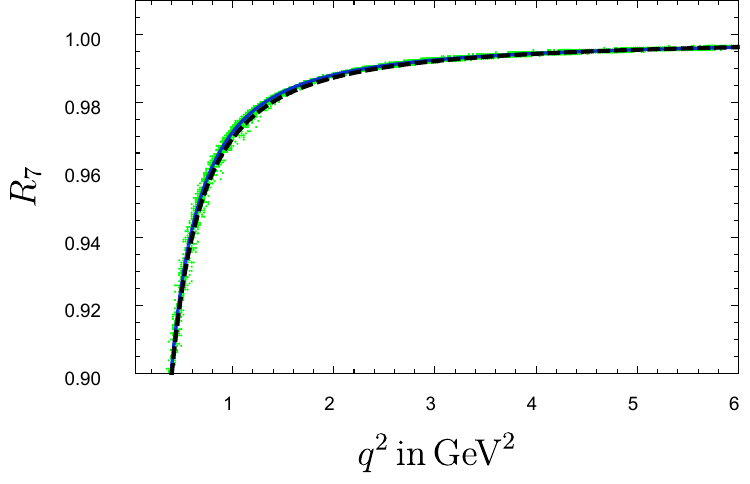} 
\includegraphics[width=0.4\textwidth]{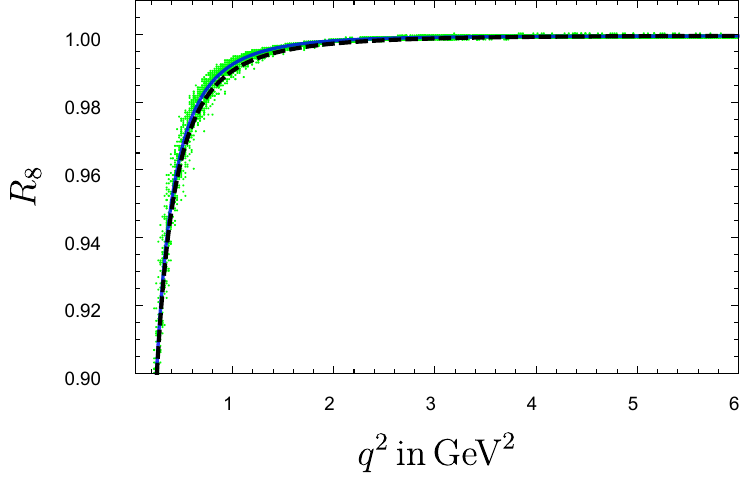} 
\caption{The $q^2$ dependence of the ratios of other observables $R_2,\, R_5,\, 
R_7,\, R_8$ that are not simple forward-backward asymmetries (See 
Eqn~\ref{R2}-\ref{R8}). The color code is the same as in Fig~\ref{Fig:3}.}.
\label{Fig:4} 
\end{figure*} 
$A_{FB}^{l}$, $A_{FB}^{h}$, $A_{FB}^{hl}$ are lepton side 
forward-backward asymmetry, hadron side forward backward asymmetry and double 
forward-backward asymmetry respectively. The parameter $\alpha_{\Lambda}$ is 
the asymmetry parameter of the decay $\Lambda\rightarrow p^+\pi^-$ which is 
defined as, 
\begin{eqnarray}
\alpha_{\Lambda}=\frac{\beta}{\tau}=\frac{\vert h_{-\frac{1}{2},0} \vert^2 - 
\vert h_{\frac{1}{2},0} \vert^2}{\vert h_{-\frac{1}{2},0} \vert^2 + \vert 
h_{\frac{1}{2},0} \vert^2}
\end{eqnarray}
Note that the convention used by us is same as in \cite{Gutsche:2013pp} upto an
overall negative sign. The asymmetry parameter has been measured to be
$\alpha_{\Lambda}=0.642\pm 0.013$ \cite{Beringer:1900zz}. As mentioned already,
this is in contrast to the mesonic counterpart $B\rightarrow K^{*}(\rightarrow
K\pi)\ell^+\ell^-$, where the subsequent $K^{*}\rightarrow K\pi$ decay is parity
conserving and thus no forward-backward asymmetry in the hadronic angle
$\theta_{K^*}$ is observed. While we are discussing  $A_{FB}^{h}$ we would also
like to point out that it is sensitive to the timelike polarization of
$j_{\eff}$ as the presence of $A_{t}$ can be seen in Eq.~\eqref{ratio2}. If we
are to restrict ourselves to the SM effective operators, the transversity
amplitude $A_{t}$ involves the Wilson coefficient $C_{10}$ only. $A_{t}$
receives additional contribution in presence of pseudoscalar operators of the
form $\left(\bar{s}\gamma_{5}b\right)(l\gamma_{5} l )$ as shown in
\cite{Altmannshofer:2008dz}. Thus, $A_{FB}^{h}$ provides an independent test of
pseudoscalar currents that are not present in the SM. The lepton
forward-backward asymmetry is given by $A^{l}_{FB}$, which depends on the real
part of the interference between two amplitudes $A_{\perp,1}$ and
$A_{\parallel,1}$. The presence of the factor $v$  suggests that lepton
forward-backward asymmetry vanishes as $q^2\rightarrow 4m_{l}^2$. 
The double forward-backward asymmetry is given by $A_{FB}^{hl}$. It is 
clear that $A_{FB}^{hl}$ vanishes when either asymmetry parameter 
$\alpha_{\Lambda}=0$ or  $q^2\rightarrow 4m_{l}^2$.

An interesting feature of the three forward-backward observables  
$A_{FB}^{l}$, $A_{FB}^{h}$ and $A_{FB}^{hl}$ is their characteristic-$q^2$ 
dependence. More precisely, within the SM one finds that both $ A_{FB}^{l}$ and 
$A_{FB}^{hl}$ cross zero, in contrast to $A_{FB}^{h}$ which doesn't. Moreover, 
to the leading order,  the zero crossing points are same for $A_{FB}^{l}$ and 
$A_{FB}^{hl}$. The $q_{0}^2$ value 
only depends on ratios of Wilson coefficients, a result well known from  other 
exclusive and inclusive $b\to s \ell^+ \ell^-$ decay, and to leading order,
\begin{eqnarray}
q^2_{0}
\approx -\frac{2m_{b}m_{\Lambda_{b}}C_{7}}{C_{9}}
\end{eqnarray}  
If we relax the assumption that  Wilson coefficients to be flavor blind and
allow for the possibility of $C_{9}^{\mu}\neq C_{9}^{e}$, then the zero  
crossing point will be different for muon and electron. Thus, by observing
$q_{0}^2$, the zero crossing values for observables like $A_{FB}^{e}$ and 
$A_{FB}^{\mu}$, one can extract vital information about the underlying flavor 
structure of the theory. There are also 
other theoretically clean observables having zero crossing 
point in large-$q^2$ region as emphasized in Ref~\cite{Kumar:2015tnz}. Thus, a 
careful study of the zero crossing points of these observables is necessary 
over the whole $q^2$ range to disentangle genuine new physics contribution, 
which is expected to be $q^2$ independent, from any unaccounted hadronic 
effects.

We construct ratios of $A_{FB}^{\mu}$, $A_{FB,\mu}^{h}$, 
$A_{FB}^{h\mu}$ to the corresponding quantities for the electron i.e. 
$A_{FB}^{e}$, $A_{FB,e}^{h}$, $A_{FB}^{he}$. These ratios are defined in a similar vein as done in ~\cite{Capdevila:2016ivx, Descotes-Genon:2015uva, Jager:2014rwa, Jager:2012uw},
\begin{align}
\label{eq:Rl2}
R_{3}=R_{A_{FB}^{l}}&=\frac{A_{FB}^{\mu}}{A_{FB}^{e}} \\
\label{eq:Rh}
R_{4}=R_{A_{FB}^{h}}&=\frac{A_{FB,\mu}^{h}}{A_{FB,e}^{h}} \\
\label{eq:Rhl}
R_{6}=R_{A_{FB}^{hl}}&=\frac{A_{FB,\mu}^{hl}}{A_{FB,e}^{hl}}.
\end{align}
We additionally provide here ratios of other angular observables for muons namely $A_2 
^{\mu}$, $A_5^{\mu}$, $A_7^{\mu}$, $A_8^{\mu}$ to the corresponding quantities 
for the electrons. 
\begin{eqnarray}
R_{2}=\frac{A_2^{\mu}}{A_2^{e}} \label{R2}\\
R_{5}=\frac{A_5^{\mu}}{A_5^{e}} \label{R5} \\
R_{7}=\frac{A_7^{\mu}}{A_7^{e}} \label{R7} \\
R_{8}=\frac{A_8^{\mu}}{A_8^{e}} \label{R8}
\end{eqnarray} 
A natural question that arises is ``to what extent the ratios defined in Eqn.\eqref{ratio1}-\eqref{ratio3} deviate from the case where individual form
factors are only known to a certain accuracy.'' In SCET the helicity amplitudes
defined in Eqn.\eqref{eq:H-F-relation} are expressible in terms of the
parameter $\xi_{\lambda}$. This simplification leads to $\xi_{\lambda}$ getting
factored out and it cancels when ratios like $A_{FB}^{l}$, $A_{FB}^{h}$,
$A_{FB}^{hl}$ are defined. In Fig.~\ref{Fig:2} we plot the $A_{FB}^{l}$,
$A_{FB}^{h}$, $A_{FB}^{hl}$ for the case of $\ell=e$ and $\ell=\mu$ separately
in the low $q^2$ region. There is however dependence on form-factor in the
observable of especial interest, the hadronic forward-backward asymmetry
$A_{FB,l}^{h}$ as can be seen in Fig.~\ref{Fig:2}. Fortunately it turns out that
the dependence on choice of form-factors cancels out in the ratios of
asymmetries defined as $R_{A_{FB}^{l}}$, $R_{A_{FB}^{h}}$ and $R_{A_{FB}^{hl}}$.
Nevertheless, if the measurement of these ratios differ from the predicted ones,
one may question the accuracy of  form-factors which are only calculated based
on a model. In order to ascertain the sensitivity of these ratios due to 
inaccuracies
in the form factors, we randomly add $\pm 30\%$ error to each form factor
estimates in covariant quark model and generate
$10^4$ points to evaluate the ratios $R_{A_{FB}^{l}}$,
$R_{A_{FB}^{h}}$ and $R_{A_{FB}^{hl}}$. In Fig.~\ref{Fig:3}, we have plotted
these ratios $R_{A_{FB}^{l}}$, $R_{A_{FB}^{l}}$  and $R_{A_{FB}^{hl}}$. In
contrast to $R_{A_{FB}^{l}}$ and $R_{A_{FB}^{hl}}$, $R_{A_{FB}^{h}}$ is a ratio
of two non-vanishing asymmetries at low-$q^2$, hence, it is likely to be more
accurately measured.
The ratio of the remaining angular observables defined in Eqn.(\ref{R2})-(\ref{R8}) are plotted in Fig.~\ref{Fig:4}.
 We conclude this section by providing a numerical estimate
of the ratios $R_{\Lambda}$, $R_{A_{FB}^{l}}$, $R_{A_{FB}^{h}}$ in Table ~\ref{Table:4} for two
$q^2$-integrated bins. All the ratios show a remarkable insensitivity to the
form factor uncertainties.
In the above analysis we have not considered the contribution from 
long-distance effects. However, this is unlikely to affect our conclusions 
based on a recent study~\cite{Blake:2017fyh} of long-distance effects in the 
analogous mode $B\to 
K^* \mu^+\mu^-$. It has been shown in Ref.~\cite{Blake:2017fyh} that $\rkst$ 
is insensitive to long distance effects within the realm of SM.

\begin{center}
\begin{table}
\begin{tabular}{c c c}
\hline
\hline
Binned Ratios &Bin 1 & Bin 2\\

 & $q^2\sim$ 0.045-1\gev$^2$  & $q^2\sim$ 1-6\gev$^2$\\
\hline
$R_{\Lambda}$&0.907$\pm$ 0.003&0.9885 $\pm$ 0.0002\\

$R_{A^{l}_{FB}}$& 0.9469 $\pm$ 0.0007 & 0.998$\pm$0.196\\

$R_{A^{h}_{FB}}$&0.993$\pm$0.001&$>0.9973$ ($0.999$\%C.L.)\\
\hline
\hline
\end{tabular}
\caption{The binned values of the observables $R_{\Lambda}$, $R_{A_{FB}^{l}}$, 
$R_{A_{FB}^{h}}$. Only $R_{A_{FB}^{h}}$ for $q^2\sim$ 1-6 \gev$^2$ does not 
show a Gaussian behavior as it is peaked towards unity.}
\label{Table:4}
\end{table}
\end{center}

\section{Conclusions} The baryonic decay mode $\Lambda_{b}\rightarrow \Lambda
\ell^{+}\ell^{-}$ is similar at the quark level to the much studied mesonic
decay mode $B\to\kstar\ell^+\ell^-$ and is hence, also expected to provide a
plethora of observables that can be used to probe NP and better understand the
hadronic effects accompanying the weak decay.  While this mode has been a
subject of several studies, we have reexamined the decay mode with focus on
aspects that have not been studied in detail earlier. 
We have derived the angular distribution without any approximations. In
particular, we retain the finite lepton mass effects and the two time like
amplitudes. These contributions play a significant role in estimating accurately
the size of lepton non-universality that may show up in the mode within SM.  We
estimate $\rl$ which is defined in a manner identical to $\rkrkst$ [see
Eqs.~\eqref{eq:rkrkst} and \eqref{eq:rl}]. 
The non-zero lepton mass effects become increasingly important in low-$q^2$
region where the discrepancy between  SM expectation and the experimentally
observed value of $\rkrkst$ is largest. The observation of a similar
discrepancy in $R_{\Lambda}$ is therefore necessary to substantiate the idea of
lepton non-universality in FCNC processes since such observations cannot be
restricted to the $B\rightarrow K^{(*)}\ell^{+}\ell^{-}$ alone. A discrepancy
between the estimates presented here and upcoming measurements at LHCb, would
establish that the existence of non-universality in interactions involving
fermions  on firm footing.

The angular distribution of the decay products in $\Lambda_{b}\rightarrow
\Lambda(\rightarrow p^{+}\pi^{-}) \ell^{+}\ell^{-}$ decay, provides a wealth of
information on the nature of decay. This is characterized by the angular
observables expressed in terms of  helicity amplitudes. We have presented a
complete set of ten angular observables that can be measured using this decay
mode.  We study in detail the three forward-backward asymmetries $A_{FB}^{l}$,
$A_{FB}^{h}$ and $A_{FB}^{hl}$. It may be noted that no hadron angle
forward-back asymmetry exists for the mode $B\to\kstar\ell^+\ell^-$. The
asymmetry $A_{FB}^{h}$ is found to be especially interesting since it is
non-vanishing in the large recoil limit, unlike the case with
$B\to\kstar\ell^+\ell^-$ decay mode, where all asymmetries vanish in the
low-$q^2$ limit. This is a consequence of the parity violating nature of the
subsequent $\Lambda\to p\pi$ decay.  The non-vanishing asymmetry is particularly
sensitive to new physics that violates lepton flavor universality, since it
involves comparing two finite quantities for the cases of $\ell=\mu$ and
$\ell=e$ respectively. It may be noted that all the asymmetries in $B\to
K^*\ell^+\ell^-$ vanish in the low-$q^2$ limit and as such result in comparisons
between two vanishing quantities.


We numerically estimate the three asymmetries $A_{FB}^{l}$, $A_{FB}^{h}$ and
$A_{FB}^{hl}$ and the ratios $\rl$,  $R_{A_{FB}^{l}}$, $R_{A_{FB}^{h}}$,
$R_{A_{FB}^{hl}}$. In order to ascertain that our results are not very sensitive
to the choice of form-factors we use two different approaches to form factors.
We have used both the covariant quark model and soft-collinear effective theory
to calculate all the observables. We find that all above mentioned ratios are
remarkably insensitive to the choice of factors as can be seen from
Fig.~\ref{Fig:1} and Fig.~\ref{Fig:4}. In order to probe the reliability of the
values estimated for these ratios we have randomly added $\pm 30\%$ error to
each form factor estimates in covariant quark model to evaluate these ratios
as well. In contrast to $R_{A_{FB}^{l}}$ and $R_{A_{FB}^{hl}}$, $R_{A_{FB}^{h}}$
is a ratio of two non-vanishing asymmetries at low-$q^2$, hence, it is likely to
be more accurately measured. We have numerically estimated the ratios
$R_{\Lambda}$, $R_{A_{FB}^{l}}$, $R_{A_{FB}^{h}}$ in two $q^2$-integrated bins.
All the ratios show a remarkable insensitivity to the form factor uncertainties.
Since these ratios are expected to be insensitive~\cite{Blake:2017fyh} to
long-distance contributions, we conclude that, $\rl$ and 
$R_{A_{FB}^{h}}$ 
are both
experimentally and theoretically reliable observables to test new physics beyond
the standard model.



\begin{thebibliography}{99}

\bibitem{Das:2012kz} 
  D.~Das and R.~Sinha,
  Phys.\ Rev.\ D {\bf 86}, 056006 (2012)
  [arXiv:1205.1438 [hep-ph]].

\bibitem{Mandal:2014kma} 
  R.~Mandal, R.~Sinha and D.~Das,
  Phys.\ Rev.\ D {\bf 90}, no. 9, 096006 (2014)
  [arXiv:1409.3088 [hep-ph]].
  
\bibitem{Mandal:2015bsa} 
  R.~Mandal and R.~Sinha,
  Phys.\ Rev.\ D {\bf 95}, no. 1, 014026 (2017)
  [arXiv:1506.04535 [hep-ph]].
  
  
\bibitem{Kruger:2005ep} 
  F.~Kruger and J.~Matias,
  Phys.\ Rev.\ D {\bf 71}, 094009 (2005)
  [hep-ph/0502060].
  
\bibitem{Altmannshofer:2008dz} 
  W.~Altmannshofer, P.~Ball, A.~Bharucha, A.~J.~Buras, D.~M.~Straub and M.~Wick,
  JHEP {\bf 0901}, 019 (2009)
  [arXiv:0811.1214 [hep-ph]].
  
\bibitem{Bobeth:2008ij} 
  C.~Bobeth, G.~Hiller and G.~Piranishvili,
  JHEP {\bf 0807}, 106 (2008)
  [arXiv:0805.2525 [hep-ph]].
  
\bibitem{Egede:2008uy} 
  U.~Egede, T.~Hurth, J.~Matias, M.~Ramon and W.~Reece,
  JHEP {\bf 0811}, 032 (2008)
  [arXiv:0807.2589 [hep-ph]].
  
\bibitem{Bobeth:2010wg} 
  C.~Bobeth, G.~Hiller and D.~van Dyk,
  JHEP {\bf 1007}, 098 (2010)
  [arXiv:1006.5013 [hep-ph]].
  
\bibitem{Becirevic:2011bp} 
  D.~Becirevic and E.~Schneider,
  Nucl.\ Phys.\ B {\bf 854}, 321 (2012)
  [arXiv:1106.3283 [hep-ph]].
  
\bibitem{Bobeth:2012vn} 
  C.~Bobeth, G.~Hiller and D.~van Dyk,
  Phys.\ Rev.\ D {\bf 87}, no. 3, 034016 (2013)
  [Phys.\ Rev.\ D {\bf 87}, 034016 (2013)]
  [arXiv:1212.2321 [hep-ph]].
\bibitem{Hiller:2013cza} 
  G.~Hiller and R.~Zwicky,
  JHEP {\bf 1403}, 042 (2014)
  [arXiv:1312.1923 [hep-ph]].
 
\bibitem{Gratrex:2015hna} 
  J.~Gratrex, M.~Hopfer and R.~Zwicky,
  Phys.\ Rev.\ D {\bf 93}, no. 5, 054008 (2016)
  [arXiv:1506.03970 [hep-ph]].

\bibitem{Grinstein:2004vb} 
  B.~Grinstein and D.~Pirjol,
  Phys.\ Rev.\ D {\bf 70}, 114005 (2004)
  [hep-ph/0404250].
  
\bibitem{Altmannshofer:2013foa} 
  W.~Altmannshofer and D.~M.~Straub,
  Eur.\ Phys.\ J.\ C {\bf 73}, 2646 (2013)
  [arXiv:1308.1501 [hep-ph]].
  
\bibitem{Aaij:2013qta} 
  R.~Aaij {\it et al.} [LHCb Collaboration],
  Phys.\ Rev.\ Lett.\  {\bf 111}, 191801 (2013)
  [arXiv:1308.1707 [hep-ex]].

\bibitem{Chatrchyan:2013cda} 
  S.~Chatrchyan {\it et al.} [CMS Collaboration],
  Phys.\ Lett.\ B {\bf 727}, 77 (2013)
  [arXiv:1308.3409 [hep-ex]].

\bibitem{other-signals} Other observed signals of lepton non-universality are a 
consequence of $\tau$ versus $\ell=\{\mu,e\}$ non-universality given by the 
measurement of  $R(D^{(*)})=\frac{ {\rm BR}(B\to D^{(*)}\tau\nu)}{ {\rm 
BR}(B\to D^{(*)}\ell \nu)}$. Very recently LHCb has reported similar 
discrepancy in $R(J\!/\!\psi)=\frac{ {\rm BR}(B\to J\!/\!\psi\tau\nu)}{ {\rm 
BR}(B\to J\!/\!\psi\ell \nu)}$[arXiv:1711.05623[hep-ex]].  

\bibitem{Gutsche:2013oea} 
  T.~Gutsche, M.~A.~Ivanov, J.~G.~K\"orner, V.~E.~Lyubovitskij and 
  P.~Santorelli,
  Phys.\ Rev.\ D {\bf 88}, no. 11, 114018 (2013)
  [arXiv:1309.7879 [hep-ph]].
  
  
 \bibitem{Gutsche:2013pp} 
  T.~Gutsche, M.~A.~Ivanov, J.~G.~K\"orner, V.~E.~Lyubovitskij and 
  P.~Santorelli,
  Phys.\ Rev.\ D {\bf 87}, 074031 (2013)
  [arXiv:1301.3737 [hep-ph]].
 \bibitem{Boer:2014kda} 
  P.~B\"oer, T.~Feldmann and D.~van Dyk,
  JHEP {\bf 1501}, 155 (2015)
  [arXiv:1410.2115 [hep-ph]].
 \bibitem{Mott:2011cx} 
  L.~Mott and W.~Roberts,
  Int.\ J.\ Mod.\ Phys.\ A {\bf 27}, 1250016 (2012)
  [arXiv:1108.6129 [nucl-th]].
  \bibitem{Kumar:2015tnz} 
  G.~Kumar and N.~Mahajan,
  arXiv:1511.00935 [hep-ph].  
   
  
 \bibitem{Leitner:2006nb} 
  O.~Leitner, Z.~J.~Ajaltouni and E.~Conte,
  hep-ph/0602043.  
  
  
 \bibitem{Chen:2001zc}
  C.~H.~Chen and C.~Q.~Geng,
  Phys.\ Rev.\ D {\bf 64} (2001) 074001
  [hep-ph/0106193].
  
\bibitem{Huang:1998ek} 
  C.~S.~Huang and H.~G.~Yan,
  Phys.\ Rev.\ D {\bf 59}, 114022 (1999)
  Erratum: [Phys.\ Rev.\ D {\bf 61}, 039901 (2000)]
  [hep-ph/9811303].
\bibitem{Chen:2001ki} 
  C.~H.~Chen and C.~Q.~Geng,
  Phys.\ Rev.\ D {\bf 63}, 114024 (2001)
  [hep-ph/0101171].
  
\bibitem{Chen:2001sj} 
  C.~H.~Chen and C.~Q.~Geng,
  Phys.\ Lett.\ B {\bf 516}, 327 (2001)
  [hep-ph/0101201].
  
\bibitem{Aliev:2002hj} 
  T.~M.~Aliev, A.~Ozpineci and M.~Savci,
  Phys.\ Rev.\ D {\bf 65}, 115002 (2002)
  [hep-ph/0203045].
  
\bibitem{Aliev:2002nv} 
  T.~M.~Aliev, A.~Ozpineci, M.~Savci and C.~Yuce,
  Phys.\ Lett.\ B {\bf 542}, 229 (2002)
  [hep-ph/0206014].
  
\bibitem{Aliev:2002tr} 
  T.~M.~Aliev, A.~Ozpineci and M.~Savci,
  Phys.\ Rev.\ D {\bf 67}, 035007 (2003)
  [hep-ph/0211447].
  
\bibitem{Aliev:2004yf} 
  T.~M.~Aliev, V.~Bashiry and M.~Savci,
  Eur.\ Phys.\ J.\ C {\bf 38}, 283 (2004)
  [hep-ph/0409275].
  
\bibitem{Giri:2005yt} 
  A.~K.~Giri and R.~Mohanta,
  J.\ Phys.\ G {\bf 31}, 1559 (2005).
\bibitem{Aliev:2006xd} 
  T.~M.~Aliev and M.~Savci,
  Eur.\ Phys.\ J.\ C {\bf 50}, 91 (2007)
  [hep-ph/0606225].
  
\bibitem{Aliev:2006gv} 
  T.~M.~Aliev, M.~Savci and B.~B.~Sirvanli,
  Eur.\ Phys.\ J.\ C {\bf 52}, 375 (2007)
  [hep-ph/0608143].
    
\bibitem{Zolfagharpour:2007eh} 
  F.~Zolfagharpour and V.~Bashiry,
  Nucl.\ Phys.\ B {\bf 796}, 294 (2008)
  [arXiv:0707.4337 [hep-ph]].
  
\bibitem{Aslam:2008hp} 
  M.~J.~Aslam, Y.~M.~Wang and C.~D.~Lu,
  Phys.\ Rev.\ D {\bf 78}, 114032 (2008)
  [arXiv:0808.2113 [hep-ph]].
 
   \bibitem{Wang:2008sm} 
   Y.~M.~Wang, Y.~Li and C.~D.~Lu,
   Eur.\ Phys.\ J.\ C {\bf 59}, 861 (2009)
   [arXiv:0804.0648 [hep-ph]].
  
\bibitem{Wang:2009hra} 
    Y.~M.~Wang, Y.~L.~Shen and C.~D.~Lu,
    Phys.\ Rev.\ D {\bf 80}, 074012 (2009)
    [arXiv:0907.4008 [hep-ph]].

  
\bibitem{Aliev:2010uy} 
  T.~M.~Aliev, K.~Azizi and M.~Savci,
  Phys.\ Rev.\ D {\bf 81}, 056006 (2010)
  [arXiv:1001.0227 [hep-ph]].
  
\bibitem{Sahoo:2009zz} 
  S.~Sahoo, C.~K.~Das and L.~Maharana,
  Int.\ J.\ Mod.\ Phys.\ A {\bf 24}, 6223 (2009)
  [arXiv:1112.4563 [hep-ph]].

\bibitem{Wang:2015ndk} 
  Y.~M.~Wang and Y.~L.~Shen,
  JHEP {\bf 1602}, 179 (2016)
  [arXiv:1511.09036 [hep-ph]].
  
  

\bibitem{Gutsche:2017wag} 
  T.~Gutsche, M.~A.~Ivanov, J.~G.~K�rner, V.~E.~Lyubovitskij, V.~V.~Lyubushkin 
  and P.~Santorelli,
  Phys.\ Rev.\ D {\bf 96}, no. 1, 013003 (2017)
  [arXiv:1705.07299 [hep-ph]].
 
\bibitem{Detmold:2016pkz} 
  W.~Detmold and S.~Meinel,
  Phys.\ Rev.\ D {\bf 93}, no. 7, 074501 (2016)
  [arXiv:1602.01399 [hep-lat]].

\bibitem{Meinel:2016grj} 
  S.~Meinel and D.~van Dyk,
  Phys.\ Rev.\ D {\bf 94}, no. 1, 013007 (2016)
  [arXiv:1603.02974 [hep-ph]].
  
\bibitem{Faustov:2017wbh} 
  R.~N.~Faustov and V.~O.~Galkin,
  Phys.\ Rev.\ D {\bf 96}, no. 5, 053006 (2017)
  [arXiv:1705.07741 [hep-ph]].
  
\bibitem{Blake:2017une} 
  T.~Blake and M.~Kreps,
  JHEP {\bf 1711}, 138 (2017)
  doi:10.1007/JHEP11(2017)138
  [arXiv:1710.00746 [hep-ph]].  
  \bibitem{Aaij:2017vbb} 
  R.~Aaij {\it et al.} [LHCb Collaboration],
  JHEP {\bf 1708}, 055 (2017)
  [arXiv:1705.05802 [hep-ex]].
  
  
  \bibitem{bifani}
S.~Bifani (on behalf of the LHCb Collab.), CERN seminar on April 18, 2017.


  \bibitem{1406.6482}
R.~Aaij {\it et al.} [LHCb Collab.],
Phys.\ Rev.\ Lett.\  {\bf 113}, 151601 (2014).

\bibitem{sm-pred}
G.~Hiller and F.~Kruger,
Phys.\ Rev.\ D {\bf 69}, 074020 (2004);

\bibitem{Kadeer:2005aq} 
  A.~Kadeer, J.~G.~Korner and U.~Moosbrugger,
  Eur.\ Phys.\ J.\ C {\bf 59}, 27 (2009)
  [hep-ph/0511019].
 
\bibitem{Gutsche:2015mxa} 
  T.~Gutsche, M.~A.~Ivanov, J.~G.~Körner, V.~E.~Lyubovitskij, P.~Santorelli and N.~Habyl,
  Phys.\ Rev.\ D {\bf 91}, no. 7, 074001 (2015)
  Erratum: [Phys.\ Rev.\ D {\bf 91}, no. 11, 119907 (2015)]
  [arXiv:1502.04864 [hep-ph]].
  
  
  
  



\bibitem{Chung:1971ri} 
  S.~U.~Chung,
  CERN-71-08.
\bibitem{Bialas:1992ny} 
  P.~Bialas, J.~G.~K\"orner, M.~Kramer and K.~Zalewski,
  Z.\ Phys.\ C {\bf 57}, 115 (1993).
  
\bibitem{Richman:1984gh} 
  J.~D.~Richman,
  CALT-68-1148.
  
\bibitem{Korner:2014bca} 
  J.~G.~K\"orner,
  arXiv:1402.2787 [hep-ph].
  
 \bibitem{Buchalla:1995vs} 
  G.~Buchalla, A.~J.~Buras and M.~E.~Lautenbacher,
  Rev.\ Mod.\ Phys.\  {\bf 68}, 1125 (1996)
  [hep-ph/9512380].


\bibitem{Korner:1992uw} 
  J.~G.~K\"orner and P.~Kroll,
  Z.\ Phys.\ C {\bf 57}, 383 (1993).
  
\bibitem{Korner:1991zx} 
  J.~G.~K\"orner and P.~Kroll,
  Phys.\ Lett.\ B {\bf 293}, 201 (1992).

\bibitem{Isgur:1990jg} 
  N.~Isgur,
  Phys.\ Rev.\ D {\bf 43}, 810 (1991).

\bibitem{Dosch:1997zx} 
  H.~G.~Dosch, E.~Ferreira, M.~Nielsen and R.~Rosenfeld,
  Phys.\ Lett.\ B {\bf 431}, 173 (1998)
  [hep-ph/9712350].

\bibitem{Mannel:2011xg} 
  T.~Mannel and Y.~M.~Wang,
  JHEP {\bf 1112}, 067 (2011)
  [arXiv:1111.1849 [hep-ph]].
  
  
 \bibitem{Wang:2011uv} 
  W.~Wang,
  Phys.\ Lett.\ B {\bf 708}, 119 (2012)
  [arXiv:1112.0237 [hep-ph]].
  
  


  \bibitem{Feldmann:2011xf} 
  T.~Feldmann and M.~W.~Y.~Yip,
  Phys.\ Rev.\ D {\bf 85}, 014035 (2012)
  Erratum: [Phys.\ Rev.\ D {\bf 86}, 079901 (2012)]
  [arXiv:1111.1844 [hep-ph]].
  

  
\bibitem{Beringer:1900zz} 
  J.~Beringer {\it et al.} [Particle Data Group],
  Phys.\ Rev.\ D {\bf 86}, 010001 (2012).
   
\bibitem{Capdevila:2016ivx} 
  B.~Capdevila, S.~Descotes-Genon, J.~Matias and J.~Virto,
  JHEP {\bf 1610}, 075 (2016)
  [arXiv:1605.03156 [hep-ph]].
\bibitem{Descotes-Genon:2015uva} 
  S.~Descotes-Genon, L.~Hofer, J.~Matias and J.~Virto,
  JHEP {\bf 1606}, 092 (2016)
  [arXiv:1510.04239 [hep-ph]].

\bibitem{Jager:2014rwa} 
  S.~J\"ager and J.~Martin Camalich,
  Phys.\ Rev.\ D {\bf 93}, no. 1, 014028 (2016)
  [arXiv:1412.3183 [hep-ph]].
  
\bibitem{Jager:2012uw} 
  S.~J\"ager and J.~Martin Camalich,
  JHEP {\bf 1305}, 043 (2013)
  [arXiv:1212.2263 [hep-ph]].
     
\bibitem{Blake:2017fyh} 
  T.~Blake, U.~Egede, P.~Owen, G.~Pomery and K.~A.~Petridis,
  arXiv:1709.03921 [hep-ph].

  


\end{thebibliography}
\end{document}